\documentclass[twocolumn, notitlepage, secnumarabic,amssymb, aps, prb, figure, superscriptaddress, floatfix] {revtex4-1}

\usepackage{natbib} 
\usepackage{balance}
\usepackage[T1]{fontenc}
\usepackage[compact]{titlesec}
\usepackage{graphicx}
\usepackage{amsmath}
\usepackage{subfigure}
\usepackage{braket}
\usepackage{hyperref}
\usepackage{bm}
\hypersetup{
    colorlinks=true,
    linkcolor=blue,
    filecolor=red,      
    urlcolor=blue,
    citecolor=black
}
\usepackage[version=4]{mhchem}
\usepackage{ragged2e}
\usepackage{float}

\usepackage[usenames, dvipsnames]{color}
\definecolor{mypink}{RGB}{219, 48, 122}
\usepackage[pagewise]{lineno}
\begin{document}

\title{Active Sparse Bayesian Committee Machine Potential for Isothermal-Isobaric Molecular Dynamics Simulations}

\author{Soohaeng Yoo Willow}
\affiliation{Department of Energy Science, Sungkyunkwan University, Seobu-ro 2066, Suwon, 16419, Korea}

\author{Dong Geon Kim}
\affiliation{Department of Energy Science, Sungkyunkwan University, Seobu-ro 2066, Suwon, 16419, Korea}

\author{R. Sundheep}
\affiliation{Department of Energy Science, Sungkyunkwan University, Seobu-ro 2066, Suwon, 16419, Korea}

\author{Amir Hajibabaei}
\affiliation{Yusuf Hamied Department of Chemistry, University of Cambridge, Lensﬁeld Road, Cambridge, CB2 1EW, United Kingdom}

\author{Kwang S. Kim}
\affiliation{Center for Superfunctional Materials, Department of Chemistry, Ulsan National Institute of Science and Technology,  Ulsan 44919, Korea}  

\author{Chang Woo Myung}
\email{cwmyung@skku.edu}
\affiliation{Department of Energy Science, Sungkyunkwan University, Seobu-ro 2066, Suwon, 16419, Korea}

\date{\today}

\begin{abstract}
Recent advancements in machine learning potentials (MLPs) have significantly impacted the fields of chemistry, physics, and biology by enabling large-scale first-principles simulations. Among different machine learning approaches, kernel-based MLPs distinguish themselves through their ability to handle small datasets, quantify uncertainties, and minimize over-fitting. Nevertheless, their extensive computational requirements present considerable challenges. To alleviate these, sparsification methods have been developed, aiming to reduce computational scaling without compromising accuracy. In the context of isothermal and isobaric ML molecular dynamics (MD) simulations, achieving precise pressure estimation is crucial for reproducing reliable system behavior under constant pressure. Despite progress, sparse kernel MLPs struggle with precise pressure prediction. Here, we introduce a virial kernel function that significantly enhances pressure estimation accuracy of MLPs. Additionally, we propose the active sparse Bayesian committee machine (BCM) potential, an on-the-fly MLP architecture that aggregates local sparse Gaussian process regression (SGPR) MLPs. The sparse BCM potential overcomes the steep computational scaling with the kernel size, and a predefined restriction on the size of kernel allows for a fast and efficient on-the-fly training. Our advancements facilitate accurate and computationally efficient machine learning-enhanced MD (MLMD) simulations across diverse systems, including ice-liquid coexisting phases, \ce{Li10Ge(PS6)2} lithium solid electrolyte, and high-pressure liquid boron nitride.
\end{abstract}

\maketitle

\section{Introduction}
In the last two decades, the application of machine learning  potentials (MLPs) in the field of materials simulations has proven to be highly successful,  particularly for the accurate prediction of potential energy surfaces (PESs) of the reference ab initio calculations,\cite{unkeMachineLearningForce2021,DeringerCsanyi21,BartokCsanyi10,ChenMarkland23a,gilmerNeuralMessagePassing2017,batznerEquivariantGraphNeural2022,hajibabaeiSparseGaussianProcess2021,hajibabaeiUniversalMachineLearning2021,HajibabaeiKim21b,myungChallengesOpportunitiesProspects2022,VandermauseKozinsky22,VandermauseKozinsky20,metcalfApproachesMachineLearning2020,naReverseGraphSelfattention2021,zhung3DMolecularGenerative2024,imbalzanoUncertaintyEstimationMolecular2021,Bayerl2022} such as density functional theory (DFT),\cite{laasonenInitioLiquidWater1993, richardsDesignSynthesisSuperionic2016, boeroFirstPrinciplesMolecular1998}  many-body perturbation (MP),\cite{WillowHirata15,delbenBulkLiquidWater2013,delbenForcesStressSecond2015} and coupled-cluster (CC) theories.
Moreover, MLPs have an affordable computational cost, allowing one to investigate previously inaccessible chemical phenomena at the first-principles level with currently accessible supercomputers. 
This breakthrough has paved the way for large-scale and long-time simulations in various scientific domains, including physics, chemistry, biology, and materials science.
Several distinct MLP methods have been developed, such as the Behler-Parrinello MLP,\cite{behlerGeneralizedNeuralNetworkRepresentation2007,behlerAtomcenteredSymmetryFunctions2011,punPhysicallyInformedArtificial2019,eckhoffHighdimensionalNeuralNetwork2021,schranCommitteeNeuralNetwork2020,schranMachineLearningPotentials2021,kapilFirstprinciplesPhaseDiagram2022} Gaussian approximation potentials (GAPs),\cite{BartokCsanyi10,bartokRepresentingChemicalEnvironments2013,bartokAussianApproximationPotentials2015,klawohnGaussianApproximationPotentials2023}
sparse Gaussian process regression (SGPR).\cite{hajibabaeiSparseGaussianProcess2021,hajibabaeiUniversalMachineLearning2021,quinonero-candelaUnifyingViewSparse2005,HajibabaeiKim21b,myungChallengesOpportunitiesProspects2022}, graph neural network potentials (GNNs),\cite{gilmerNeuralMessagePassing2017,batznerEquivariantGraphNeural2022} gradient-domain machine learning (GDML),\cite{chmielaMachineLearningAccurate2017} moment tensor potentials (MTP),\cite{shapeevMomentTensorPotentials2016,podryabinkinActiveLearningLinearly2017,NovikovShapeev21} and equivariant message-passing neural networks.\cite{batznerEquivariantGraphNeural2022,NEURIPS2022_4a36c3c5}
The versatility of these methodologies has led to their widespread adoption and successful application across different scientific disciplines, promising valuable insights and outcomes.\cite{lim_mlsolva_2021}

The Gaussian Process ($\mathcal{GP}$) based MLP boasts in the active learning, uncertainty prediction, and low data requirement.\cite{rasmussenGaussianProcessesMachine2005,shahriariTakingHumanOut2016,lawrenceProbabilisticNonlinearPrincipal2005,alvarezComputationallyEfficientConvolved2011}
However, despite all these advantages, the computational scalability of $\mathcal{GP}$-based MLPs comes under scrutiny. 
$\mathcal{GP}$-based MLPs face computational challenges when dealing with a large dataset, whose training complexity increases dramatically to $\mathcal{O}(n^3)$ with the size of kernel matrix $n$ 
due to its inversion. As a result, $\mathcal{GP}$s are restricted to handling the datasets whose size are smaller than $10^4$. 
This limitation led to the development of sparsification techniques to select a subset of data point $m$ among $n$ data points to reduce the computational complexity.\cite{hajibabaeiSparseGaussianProcess2021,hajibabaeiUniversalMachineLearning2021,quinonero-candelaUnifyingViewSparse2005,HajibabaeiKim21b,myungChallengesOpportunitiesProspects2022}
However, the sparsification approaches including SGPR still struggles with large number of inducing sets $m$ as $\mathcal{O}(nm^2)$ for training. 
Another strategy to mitigate the steep computational scaling is to use aggregation models, including the Bayesian Committee Machine (BCM)\cite{trespBayesianCommitteeMachine2000}. The BCM, for instance, combines the predictions from $P$ distinct local expert models, each trained on different sub-data. This further reduces the computational scaling to $\mathcal{O}(nm^2/P^3)$ for training.  

MD simulations serve as a powerful tool to complement experimental findings by offering a detailed view of the physical system at the atomic scale. 
By comparing the results of MD simulations with their corresponding experimental observations (i.e., phase diagrams, reaction kinetics, and structure factor), one can effectively test the accuracy and reliability of the given PES.
To perform isobaric simulations, such as isothermal-isobaric ($NpT$) and isobaric-isenthalpic ($NpH$) MD simulations, it is crucial to accurately estimate the pressure of system.
In this regard, improving the accuracy of pressure estimation of MLPs could enhance the reliability of isobaric ML MD simulations.
While other types of MLPs such as neural network MLP \cite{wangDeePMDkitDeepLearning2018} and GAPs,\cite{klawohnGaussianApproximationPotentials2023} incorporated the estimation of virial pressure, 
SGPR-based MLP has not yet included the estimation of virial pressure.\cite{hajibabaeiSparseGaussianProcess2021}

Zhang et al.\cite{zhangPhaseDiagramDeep2021} employed MD simulations with a SCAN-DFT\cite{sunStronglyConstrainedAppropriately2015} based deep potential model\cite{zhangActiveLearningUniformly2019,wangDeePMDkitDeepLearning2018} to predict ice-water phase diagram, achieving a generally satisfactory agreement with experimental results.
Kapil et al.\cite{kapilFirstprinciplesPhaseDiagram2022} predicted the pressure-temperature phase diagram of monolayer-confined water using committee neural network potentials (C-NNP)\cite{schranCommitteeNeuralNetwork2020} at the accuracy of quantum Monte Carlo (QMC).
Subsequently, Bore and Paesani in 2023\cite{BorePaesani23} utilized the deep potential model\cite{wangDeePMDkitDeepLearning2018} based on MB-pol\cite{babinDevelopmentFirstPrinciples2013,babinDevelopmentFirstPrinciples2014,rieraMBXManybodyEnergy2023} to estimate the ice-water phase diagram and showed improved agreement with experimental results.
Surprisingly, however, they observed that the phase diagram predicted by the MLP did not mirror the one obtained using MB-pol potential. And it demonstrates that well-sampled atomic configuration training data is required to build accurate MLPs. Also, it underscores the significance of accurate pressure prediction of system by MLPs. Motivated by these, we develop a virial
kernel function designed to incorporate virial prediction into the covariance matrix to enhance the accuracy of pressure predictions. 
In addition, we develop the on-the-fly active BCM potential to train MLPs based on the training dataset that prioritises the inclusion of structures with high uncertainty. As the BCM is an aggregation model, the BCM potential enhances the scalability and transferability of SGPR MLPs. 
We demonstrate these advantages of BCM potential with three different systems: ice-liquid coexisting phases (ice II-water, ice III-water, and ice V-water), \ce{Li10Ge(PS6)2} (LGPS) solid electrolyte, and high-pressure boron nitride liquid. 

\begin{figure*} [htp]
\centering
\includegraphics[width=13cm]{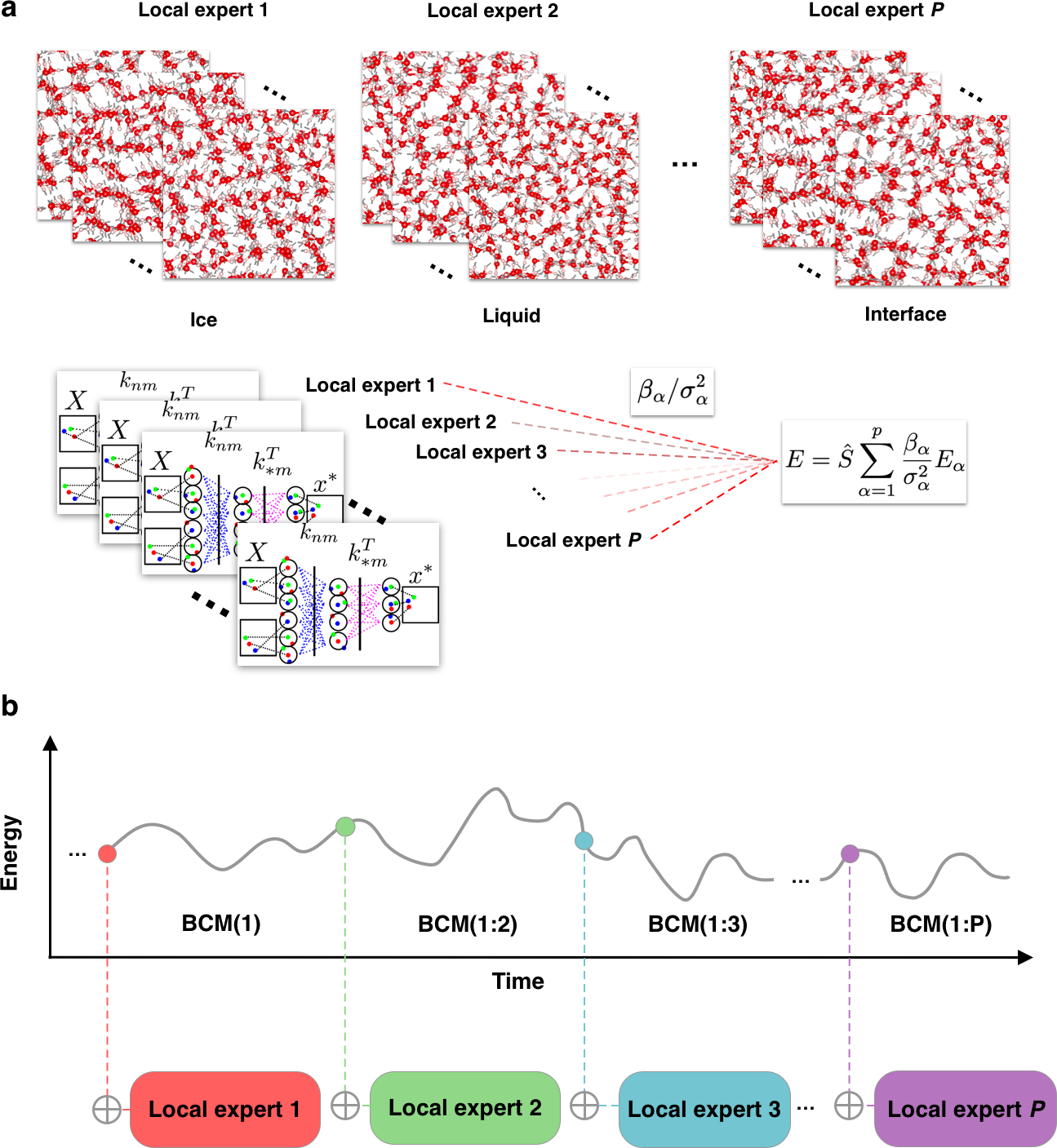}
\caption{Schematic of active sparse Bayesian committee machine (BCM) potential. (a)
The entire dataset is divided into $P$ distinct subsets. 
Within each subset, a sparse Gaussian process regression (SGPR) MLP is trained.
Subsequently, a combined version of MLP for the entire dataset is built through the BCM algorithm. The weights $\frac{\beta_\alpha}{\sigma_\alpha^2}$ ensure that the predictions from confident models are more outspoken than those from less confident models. (b) Illustration of active BCM algorithm. Each local expert SGPR model undergoes training up to a predefined kernel size. Before surpassing the kernel size limit, our algorithm initiates training for the next SGPR model. This process is sequentially executed, resulting in the assembly of $P$ local SGPR MLPs, BCM(1:$P$), forming a Bayesian committee. Throughout the MD simulation, this ensemble of $P$ committee MLPs predicts the energy, forces, and virial tensor of the system.}
\label{fig:BCP}
\end{figure*}

\section{Methods}

All MD simulations were carried out using the ASE package.\cite{larsen_atomic_2017} We employed a time step of 0.5 fs for the integration of the equations of motion that were propagated according to combined Nos{\'e}-Hoover and Parrinello-Rahman dynamics.\cite{melchionnaHooverNPTDynamics1993,melchionnaConstrainedSystemsStatistical2000,holianTimereversibleEquilibriumNonequilibrium1990} To acquire the training and testing datasets for analysis, three different systems were chosen. 
For liquid boron Nitride (BN) we trained the BCM potential to DFT at the r$^2$SCAN-D4\cite{r2scan, dftd4}, while for LGPS the potential was trained at the Perdew-Burke-Ernzerhof (PBE) level\cite{perdewRationaleMixingExact1996}.
All the DFT calculations with projector augmented wave (PAW) pseudopotentials\cite{PAW} were performed using the Vienna Ab initio  Simulation Package (VASP).\cite{PhysRevB.54.11169,kresseEfficiencyAbinitioTotal1996} We performed isothermal ($NVT$) MD simulations of liquid boron nitride (BN) at temperature of $T =$ 7500 K. The liquid BN system consists of 64 atoms within the cubic box of 7.252 \AA. We performed isothermal-isobaric ($NpT$) MD simulations of LGPS at the ambient pressure and temperature ranging from $T=$ 600 to 1200 K. 
A $\Gamma$-point sampling was employed with an energy cutoff set at 500 eV.  We sampled various structures of ice phases (II, III, and V) and liquid water through $NpT$ MD simulations using the MB-pol force field\cite{babinDevelopmentFirstPrinciples2013,babinDevelopmentFirstPrinciples2014,rieraMBXManybodyEnergy2023} at pressures of $p=$ 0.4 and 0.6 GPa, with a temperature of $T=$ 255 K. 

\section {Theory}

\subsection{Virial Kernel in Sparse Gaussian Process Regression potential \label{sec1}}
The SGPR energy is given as decomposed local energy function of a local chemical environment (LCE) $\rho_i$ as 
\begin{equation} \label{eq:mlp_enr}
    E = \sum_{i=1}^N \mathcal{E}(\rho_i),
\end{equation}
where  $\mathcal{E}(\rho_i)$ is a local energy of LCE $\rho_i$.
This indicates that the SGPR energy as well as force acting on atom $i$ depends on the relative coordinates of surrounding atoms because the LCE of atom $i$ is defined by 
\begin{equation}
    \rho_i = \{ \vec{r}_{ij}; j \neq i \quad |\vec{r}_{ij}| < r_c\},
\end{equation}
where $\vec{r}_{ij} = \vec{r}_j - \vec{r}_i$, $r_c$ is a cutoff radius for neighborhood relations. 

The core component of SGPR is building a covariance kernel
$\mathcal{K} (\rho_i, \rho_j)$ that encodes the similarity between $\rho_i$ and $\rho_j$. The kernel 
is invariant with respect to the symmetry operations that leaves the potential energy invariant, e.g. translations and rotations. 
Here, the smooth overlap of atomic positions (SOAP)\cite{bartokRepresentingChemicalEnvironments2013} is used for the descriptor of a similarity kernel between LCEs and is defined as
\begin{equation}
    \mathcal{K} (\rho_i, \rho_j) = \int d \hat{\bm{R}} ~|\int d\vec{r} \xi_i(\vec{r}) \xi_j(\hat{\bm{R}} \vec{r})|^2,
\end{equation}
where $\hat{\bm{R}}$ is the three-dimensional rotation operator, $\xi_i(\vec{r})$ is the atomic density in the neighborhood of $i$,
\begin{equation}
    \xi_i(\vec{r}) = \sum_{j\in \rho_i} e^{\alpha |\vec{r}- \vec{r}_{ij}|^2},
\end{equation}
and $\alpha$ is a hyperparameter, with $\alpha = 1$ in our calculations.

In the SGPR potential, we select a reduced set of LCEs $z = \{ \chi_j \}_{j=1}^m$, called inducing set, which are significantly distinct from each other and are sufficient statistics for the original data $X = \{R_1, \cdots, R_n\}$ to improve the scalability. It is likely that many LCE pairs in $X$ are similar: $\mathcal{K}(\rho_i, \rho_j) \approx 1$. With the inducing set of LCEs, the latent local energy of atom $i$ is expressed as 
\begin{equation}
    \mathcal{E} (\rho_i) = \sum_{j=1}^{m}  \mathcal{K}(\rho_i, \chi_j) ~w_j
\end{equation}
where $\bm{w} = [w_1, \cdots, w_m]$ are the trainable weights for the descriptors in $z$.
The potential energy for a given configuration $R$ becomes
\begin{equation} \label{eq:E}
    E(R) = \sum_i^N \mathcal{E}(\rho_i) = \sum_{i,j} \mathcal{K} (\rho_i, \chi_j)~w_j.
\end{equation}
The forces $f_i^\mu$ are obtained from 
\begin{equation}
    f_i^\mu =
    \sum_{j=1}^{m}  
    \left[- \sum_{k=1}^N \frac{\partial \mathcal{K} (\rho_k, \chi_j)}{\partial r_i^\mu} \right]w_j 
    = \sum_{j=1}^m  \dot{\mathcal{K}}_i^\mu(R, \chi_j)~w_j,
\end{equation}
where $\dot{\mathcal{K}}_i^\mu (R, \chi_j)$ is the covariance kernel for the force on atom $i$ in $\mu$-direction, $\mu = \{x, y, z\}$.

The previous SGPR potential didn't consider the pressure tensor in the training process.\cite{hajibabaeiSparseGaussianProcess2021} Due to this, the SGPR potential often couldn't accurately predict the pressure tensor of a given system. To improve this shortcoming, our strategy is to incorporate the pressure data into the training process of SGPR. The instantaneous pressure tensor $\mathcal{P}^{\mu \nu}$ is defined as
\begin{equation} \label{eq:P}
    \mathcal{P}^{\mu \nu} = N k_B \mathcal{T}^{\mu \nu}/V + \mathcal{W}^{\mu \nu}/V,
\end{equation}
where $\mathcal{T}^{\mu \nu}$ is the instantaneous temperature tensor. To predict the pressure tensor, we build the internal virial tensor $\mathcal{W}^{\mu \nu}$ as below
\begin{eqnarray} \label{eq:W}
    \mathcal{W}^{\mu \nu} & = & \sum_{i}^N f_{i}^\mu r_{i}^\nu \nonumber \\ 
    & = & \sum_{j}^m  \left[\sum_{i}^N \dot{\mathcal{K}}_i^\mu (R, \chi_j) r_i^\nu \right]~w_j.
\end{eqnarray}

The weight vector $\bm{w}$ is obtained by the fitting of $E(R)$, $f_i^\mu$, and $\mathcal{W}^{\mu \nu}$ 
to the target energies, forces, virial pressures $\bm{y}_n$ for a given dataset $X =\{R_1, \cdots, R_n\}$. 
The subsequent fitting process of SGPR is nicely reformulated as a form of linear equation with the regularization hyperparameter $\varepsilon$, \cite{hajibabaeiSparseGaussianProcess2021}
\begin{equation}\label{eq:wm}
    \begin{bmatrix} \bm{K}_{nm} \\  \varepsilon \bm{L}^T \end{bmatrix} \bm{w}_m = 
    \begin{bmatrix} \bm{y}_n \\ \bm{0} \end{bmatrix},
\end{equation}
where $\bm{L}$ is the Cholesky factor of $\bm{K}_{mm}$ ($\bm{K}_{mm} = \bm{L} \bm{L}^T$, where $\bm{L}$ is lower triangular) and $\bm{0}$ is a columnar vector of zeros with length $m$. $\varepsilon$ and other hyperparameters, which are defined in Eq. (\ref{eq:wm}), are optimized by maximizing the log likelihood of $\bm{y}_n$.

\subsection{Active Sparse Bayesian Committee Machine potential}

The SGPR algorithm still encounters difficulties in obtaining the solution $\bm{w}_m$ of Eq. (\ref{eq:wm}), which involves the inversion of a kernel matrix, 
with large size of training datasets $X$ and inducing set $z$ of LCEs. 
In order to mitigate this challenge, 
we employed  BCM\cite{trespBayesianCommitteeMachine2000} to combine SGPR MLPs which are separately trained on different subsets as a local expert (Figure \ref{fig:BCP} and S1). 
When building MLP of ice polymorphs, rather than utilizing all ice phases as a training dataset, we partitioned each ice phase into training subsets. 
Subsequently, we trained each local expert models for corresponding subset of ice phase. We combined these individually trained local expert SGPR models into a universal BCM MLPs. This approach not only streamlines the training process but also addresses the computational challenges associated with the inversion of large kernel matrices, paving a way for building an efficient and accurate universal MLPs.

For given subsets of the training dataset $\{n_1, \cdots, n_P\}$ and inducing LCEs $\{m_1, \cdots, m_P\}$, the BCM\cite{trespBayesianCommitteeMachine2000} potential energy is
\begin{eqnarray} \label{eq:E_bcm}
    E = \hat{S} \sum_{\alpha=1}^{P} \frac{\beta_\alpha}{\sigma_\alpha^2}E_\alpha,
\end{eqnarray}
where 
the potential energy $E_\alpha$ of $\alpha$-th SGPR local expert ($R_{n \in \alpha}$ and $z_\alpha = \{ \chi_j \}_{j \in \alpha})$ is given
\begin{eqnarray} \label{eq:E_alpha}
    E_\alpha (R) & = & 
    \sum_{i, (j\in m_\alpha)}  \mathcal{K}(\rho_i, \chi_j)~w_j^\alpha. 
\end{eqnarray}
where $\bm{w}^\alpha$ are the training weights of $\alpha$-th local SGPR model. 

The forces $f_i^\mu$ and internal virial $\mathcal{W}^{\mu \nu}$ of the BCM potential are obtained from 
\begin{eqnarray}
f_i^\mu & = &\hat{S} \sum_{\alpha=1}^{P} \frac{\beta_\alpha}{\sigma_\alpha^2}
f_{i,\alpha}^{\mu} \\
\mathcal{W}^{\mu \nu} & = & \hat{S} \sum_{\alpha=1}^{P} \frac{\beta_\alpha}{\sigma_\alpha^2} \mathcal{W}^{\mu \nu}_\alpha,
\end{eqnarray}
where $f_{i,\alpha}^\mu$ and $\mathcal{W}^{\mu \nu}_\alpha$ represent the forces and the internal virial of the $\alpha$-th SGPR local expert, which are given
\begin{eqnarray}
    f_{i,\alpha}^\mu & = &
    \sum_{j \in m_\alpha}  \dot{\mathcal{K}}_i^\mu(R, \chi_j)~w_j^\alpha, ~~\mathrm{and} \nonumber \\
    \mathcal{W}^{\mu \nu}_\alpha & = & \sum_i^N f_{i,\alpha}^\mu r_i^\nu. \nonumber
\end{eqnarray}

The maximum of the covariance loss for the $\alpha$-th subset LCEs, denoted as $\sigma_\alpha^2$, is used to weight the $\alpha$-th committee prediction. The $\sigma_\alpha^2$ is estimated by
\begin{equation}
    \sigma_\alpha^2 = \mathrm{max}(1 - \bm{K}_{*m}\bm{K}_{mm}^{-1}\bm{K}_{*m}^T),
\end{equation}
where $\bm{K}_{*m}$ is the covariance matrix between a test configuration $\bm{x}^*$ and $\alpha$-th inducing set $z_\alpha$.
Furthermore, we weight the prediction with $\beta_\alpha$ following the concepts of the robust BCM where the individual committee prediction is weighted by the differential entropy, 
$\beta_\alpha = -\log(\sigma_{\alpha}^2)$. In the BCM, a particular local expert which has the highest similarity with a test configuration $\bm{x}^*$ mostly contributes to the energy, forces, and pressure predictions. The final normalization factor $\hat{S}$ is thus given:
\begin{eqnarray}
 \hat{S} &= &\left[ \sum_\alpha^P \frac{\beta_\alpha}{\sigma_\alpha^2}\right]^{-1}.
\end{eqnarray}

The significant benefit of employing the SGPR-based BCM is that, in addition to the sparsification, we invert an approximate $(m/P)$-dimensional matrix instead of inverting an $m$-dimensional matrix ($m = \sum_\alpha^P m_\alpha$). Hence, the computational cost of the matrix inversion scales as $\mathcal{O}(m^3/P^3)$ rather than $\mathcal{O}(m^3)$. Compared to ``product of experts'' (POE),\cite{hajibabaeiUniversalMachineLearning2021} in which the potential energy prediction is $E \approx \frac{1}{P} \sum_{\alpha}^P E_\alpha$, 
the BCM scale factor $\beta_\alpha/\sigma_\alpha^2$ ensures that larger weights are applied for the predictions of confident models than less confident models. 


\section {Results and Discussion}

\subsection{Accurate pressure predictions of virial kernel method}

While neural network MLPs can accurately predict virial stress based solely on energies and forces, incorporating virial stress into the training set through a virial stress loss function has been shown to enhance performance by reducing the number of configurations required for training.\cite{zhang2018nips} In this work, we demonstrate that including virial stress in the training process is critical for accurate virial stress predictions in SGPR and BCM MLPs. In the SGPR MLP, the training weight parameter vector $\bm{w}_m$ relies on the covariance matrix $\bm{K}_{nm}$ and the corresponding target values $\bm{y}_n$ for energy, forces, virial pressure in Eq. (\ref{eq:wm}). 
The values of weight vector, as determined by the solution to Eq. (\ref{eq:wm}), vary depending on whether the virial pressure term is included in the covariance matrix and the corresponding target values.
In this subsection, we compare instantaneous pressure predictions using the weights obtained under two different approaches. This comparative analysis provides insights into the effect of virial kernel in covariance matrix and into the accuracy of pressure prediction. In our subsequent discussion, we will elucidate how the omission of the virial kernel, $\mathcal{W}^{\mu \nu}$, can precipitate abrupt transitions in energy or pressure during the update in active learning. Such discontinuities are problematic in $NpT$ or $NpH$ simulations, undermining its reliability. This phenomenon could raise significant concerns regarding the credibility of simulation results in the absence of virial kernel in SGPR potential.

The first scheme ({\bf{Scheme 1}}) involves building $\bm{w}_m$ solely with $K_{nm}$ and $\bm{y}_n$ that includes energies and forces. 
The second scheme ({\bf{Scheme 2}}) trains $\bm{w}_m$ by incorporating virial kernel into $\bm{K}_{nm}$ and $\bm{y}_n$ includes energies, forces, and pressures. To substantiate the predictive power of the virial kernel, we evaluated the accuracy of {\bf{Scheme 1}} and {\bf{Scheme 2}} across different phases of materials: high-density liquid phase BN at $T =$ 7500 K ($NVT$) (Figure S2), solid Li electrolyte LGPS at temperatures ranging from $T=$ 600 to 1000 K and at atmospheric pressure ($NpT$), and coexisting phase of ice III and liquid water at pressures of $p=$ 0.4 and 0.6 GPa, with a temperature of $T=$ 255 K ($NpH$). Through these systematic investigations across different systems and phases, we aim to offer a comprehensive evaluation of the virial kernel's predictive accuracy.

\begin{figure} [!h]
\centering
\includegraphics[width=8.0cm]{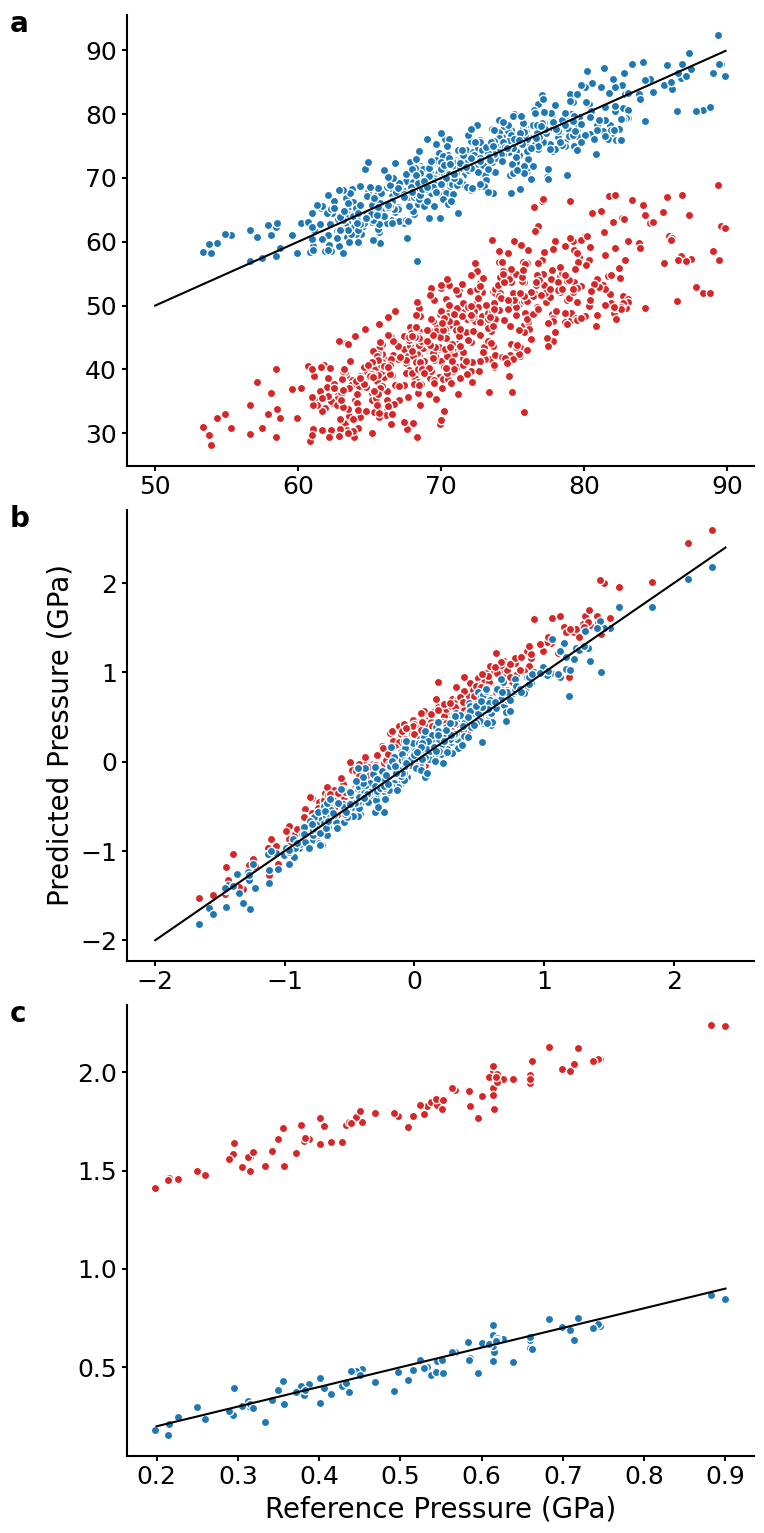}
\caption{Comparison of pressure predictions with and without virial kernel. Three different systems were tested to assess the impact of including the virial term in the covariance matrix $\bm{K}_{nm}$ on pressure calculations: (a) liquid boron nitride, (b) solid Li electrolyte LGPS, and (c) a coexisting system of ice III and liquid water.
\textbf{Scheme 1} (red dots) involves constructing $\bm{w}_m$ solely with $K_{nm}$ and target values of energies and forces. 
\textbf{Scheme 2} (blue dots) incorporates virial kernel into $\bm{K}_{nm}$ and $\bm{y}_n$ includes target energies, forces, and pressure to obtain $\bm{w}_m$. Under high pressure conditions, the predictive accuracy of \textbf{Scheme 1} for pressure exhibits significant errors.}
\label{fig:E_P}
\end{figure}

Once we trained the weight parameter vector $\bm{w}_m$, we estimated the pressures with the test dataset using Eqs. (\ref{eq:P}) and (\ref{eq:W}). This highlights that {\bf Scheme 2} is capable of accurately reproducing target instantaneous pressures, in contrast to {\bf Scheme 1} (Figure \ref{fig:E_P}). BN is an industrially important material which is manufactured at high temperature and pressure. Recently, it was proposed that amorphous BN exhibits low dielectric characteristics and can be used for high performance electronic applications.\cite{hong_ultralow-dielectric-constant_2020} Under extreme thermodynamic conditions ($T =$ 7500 K), {\bf{Scheme 1}} struggles to predict the pressure of liquid BN system, resulting in a substantially large error with a root mean square error (RMSE) of 26.4 GPa. However, the inclusion of virial kernel improves the prediction accuracy, reducing the RMSE from 26.4 GPa to 2.8 GPa (Figure \ref{fig:E_P}a).

LGPS is one of the promising candidates for solid state Li-ion electrolytes. At room temperature, LGPS maintains its crystallinity while Li ions diffuse (Figure S4). We find that the pressure predictions for LGPS from both {\bf{Scheme 1}} and {\bf{Scheme 2}} align closely with the reference values because an average predicted pressure is approximately $p=$ 0 GPa (Figure \ref{fig:E_P}b). The integration of the virial term into the covariance matrix does not significantly affect the pressure when the system is in its solid phase at ambient pressure. Still, there is a slight improvement in prediction accuracy, with an RMSE decreasing from 0.26 GPa to 0.13 GPa. 

Under the mild conditions (0.2 GPa $\leq$ $p$ $\leq$ 0.6 GPa) of the coexisting ice-liquid water system, incorporating the virial kernel significantly improved the pressure prediction from an RMSE of 1.30 GPa to 0.048 GPa (Figure \ref{fig:E_P}c). These results demonstrate that including the virial kernel function in the covariance matrix is crucial for SGPR MLP, especially when dealing with high-pressure systems.



\subsection{On-the-flying active learning $NpT$ MD simulations of LGPS}

\begin{figure} [htp]
\centering
\includegraphics[width=8.0cm]{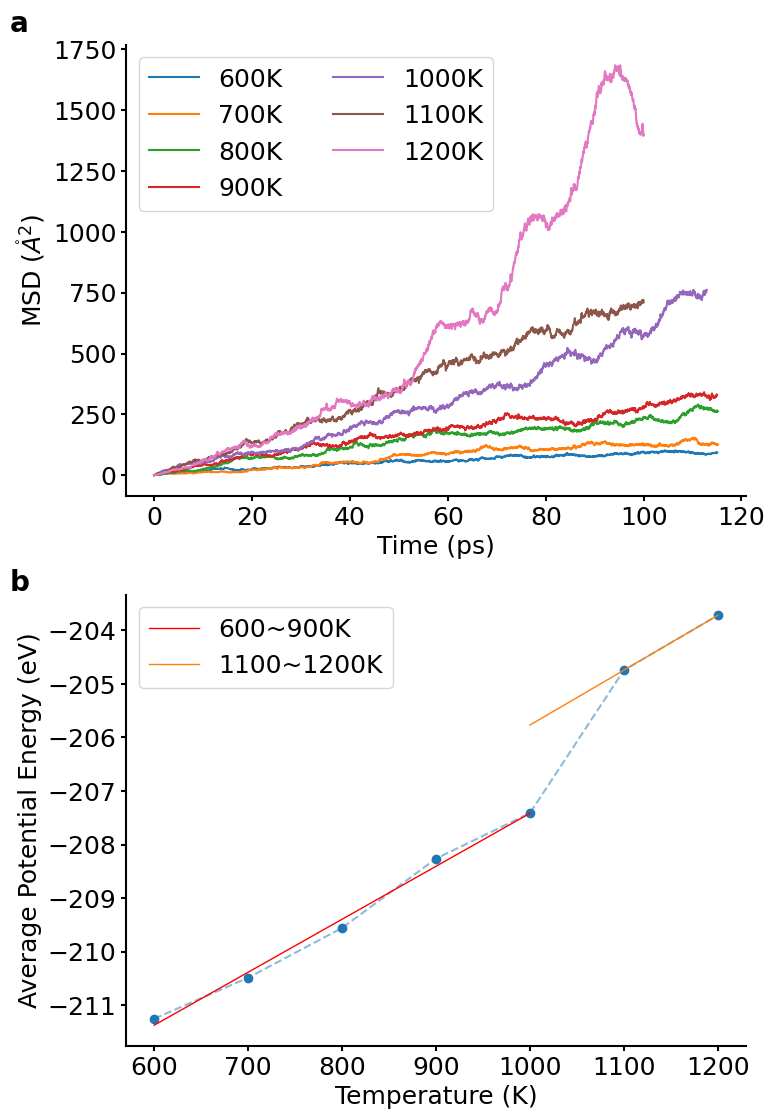}
\caption{
(a) Mean square displacement (MSD) of Li in LGPS at various temperatures (600 K: blue, 700 K: orange, 800 K: green, 900 K: red, 1000 K: purple, 1100 K: brown, 1200 K: pink). (b) Average potential energy of LGPS during $NpT$ simulations. The system undergoes phase transition from a solid to a liquid state above 1000 K.}
\label{fig:MSD_avg_PE}
\end{figure}

In 2012, Mo et. al. reported that $NVT$ first-principles MD simulations of LGPS, conducted from 600 K to 1500 K, indicated no breaking of P-S or Ge-S bonds.\cite{moFirstPrinciplesStudy2012} They concluded that LGPS is a promising and structurally stable material for all-solid-state batteries. However, at the extremely high temperature of 1500 K,  the system's pressure in a fixed volume is expected to be exceedingly high. To address this issue,
on-the-flying active learning $NpT$ MD simulations were performed under atmospheric pressure ($\sim$0.101 MPa) at various temperatures for solid Li electrolyte LGPS, with each run lasting over 100 ps (Figure \ref{fig:MSD_avg_PE} (a)).

The average potential energy of LGPS (Figure \ref{fig:MSD_avg_PE} (b)) shows two distinct linear regions: one for the solid phase (600$\sim$900 K) and another for the liquid phase (1100$\sim$1200 K). These observations from both the mean square displacement (MSD) of Li in LGPS and potential energy analyses provide clear evidence for a phase transition in LGPS around 1000 K, transitioning from a solid to a liquid state. The MSD demonstrates a clear temperature-dependent behavior, with higher temperatures resulting in increased Li ion mobility. At lower temperatures (600$\sim$900 K), the MSD shows a gradual increase, indicating solid-state diffusion. However, above 1000 K, the slope and magnitude of the MSD increase rapidly, especially at 1100 K and 1200 K. The rapid change in Li ion mobility is consistent with the potential energy discontinuity observed in Figure \ref{fig:MSD_avg_PE} (b), supporting a phase transition from a solid to a liquid state. This transition significantly enhances Li ion mobility, as evidenced by the steep increase in MSD values at higher temperatures.

This observation aligns well with the experimental data,\cite{horiPhaseDiagramLGPS2015}
wherein differential thermal analysis (DTA) was used to elucidate the phase diagram for the quasi-binary system between Li$_4$GeS$_4$ and Li$_3$PS$_4$. The experimental investigation revealed that LGPS exhibits an incongruent melting point at 650 $^\circ$C (923 K).
We confirmed that the activation energy $E_a$ for Li diffusion in LGPS in its solid state (below 1000 K) is about 0.18 eV. This value is similar to the previously reported value of 0.17 eV.\cite{hajibabaeiUniversalMachineLearning2021}

Previous computational studies on various materials, like MgO, have shown similar trends in potential energy changes with temperature.
These studies using the deep neural network based MLPs have found a sudden shifts in energy due to phase transition.\cite{Wisesa2023,Wisesa2023-2,andolinaTransferableMLP2023}

\subsection{On-the-flying active learning $NpT$ MD simulations of Liquid Water}

\begin{figure}[h!]
    \centering
    \includegraphics[width=0.45\textwidth]{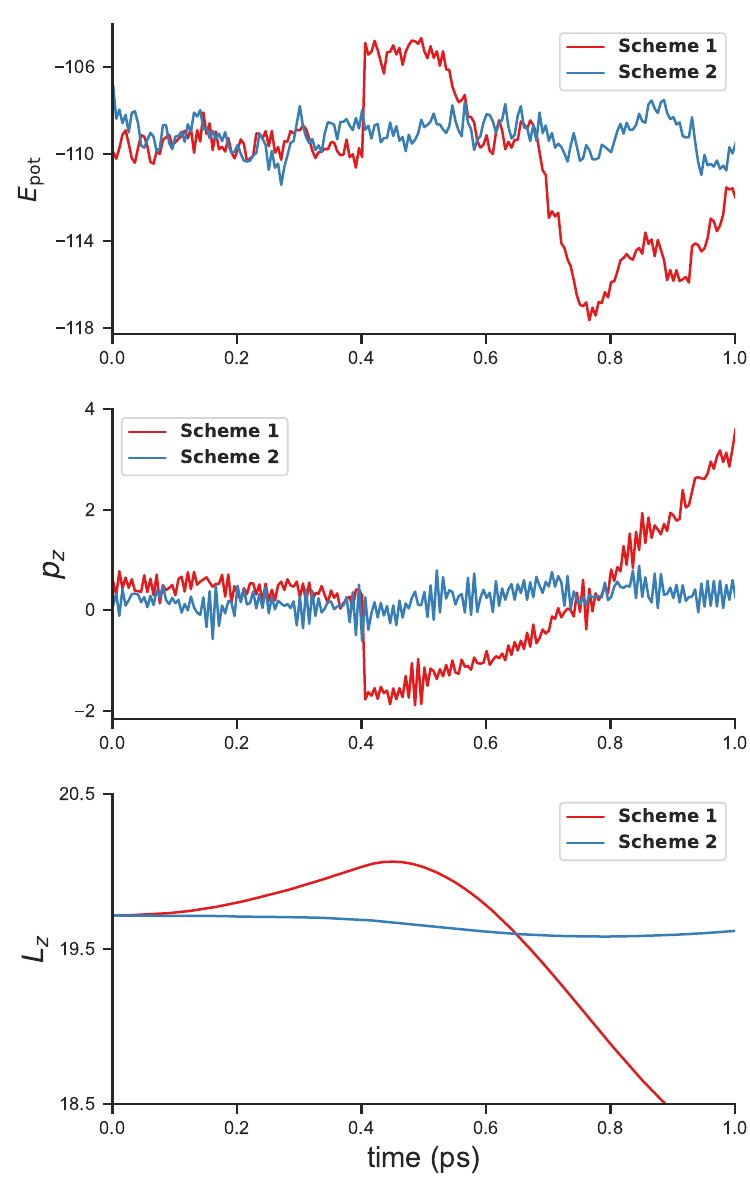}
    \caption{Fluctuations of potential energy $E_\mathrm{pot}$, pressure $p_z$, and cell $L_z$ with and without virial kernel during the active learning $NpT$ MD simulation. \textbf{Scheme 1} shows abrupt discontinuity in 
    energy (upper) and pressure (middle) according to updating dataset, inducing set, and weight vector. As a result, the $NpT$ barostat with \textbf{Scheme 1} fails for box adjustment leading to a sudden box shrinkage (lower).
    }
    \label{fig:H}
\end{figure}

On-the-flying active learning $NpT$ MD simulations were performed on liquid water under a pressure of $p=0.2$ GPa and a temperature of $T=270$ K, subsequent to training an expert model through active learning $NVT$ MD simulations at the same temperature. The initial liquid system consisted of 256 water molecules within a cubic box measuring 19.715 \AA ~in box length. Due to its inability to precisely estimate pressure,  the $NpT$ MD simulation without the inclusion of virial kernel (\textbf{Scheme 1}) failed to adjust the box's fluctuations. 
 Updating the dataset $X$, inducing set $z$, and weight vector $\bm{w}$ led to a noticeable sudden change in pressure, disrupting the isenthalipic and isobaric conditions of MD simulations (see Figure \ref{fig:H}).
On the other hands, when the virial kernel is included in the covariance matrix (\textbf{Scheme 2}), both energy and pressure predictions are smooth during the update in on-the-flying active learning $NpT$ MD simulation, resulting in precise adjustment of the box.
In this study, hence, we trained SGPR local expert MLP models for ice II, ice III, and ice V through active learning MD simulations under constant pressure and constant temperature conditions, using an active sparse BCM potential that includes the virial kernel ({\bf Scheme 2}).

\begin{figure} [htp]
\centering
\includegraphics[width=8.5cm]{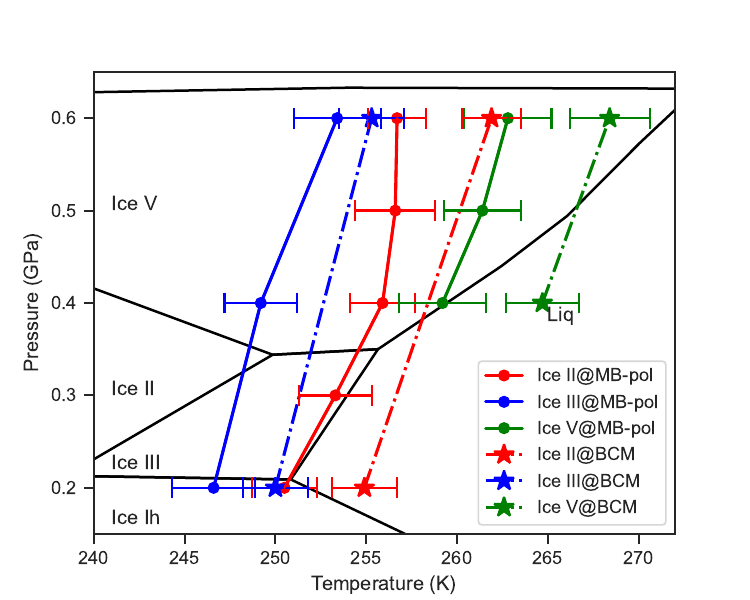}
\caption{Ice-water phase diagram.
Ice-water melting points are calculated through $NpH$ MD simulations of coexisting ice-liquid systems using the MB-pol (circle lines) and BCM (star dashed lines). 
Solid black line represents the experimental melting line.\cite{Ice_Exp}}
\label{fig:Tm}
\end{figure}

\subsection{Coexistence line of ice-liquid water}

Anisotropic $NpH$ MD simulations were employed to compute the melting temperature ($T_m$) of ice within the coexisting ice-liquid phase. 
$NpH$ MD simulations enable the spontaneous adjustment of temperature to satisfy the condition $\mu_\mathrm{ice} (p, T)_{T=T_m} = \mu_\mathrm{liq}(p,T)_{T=T_m}$. 
Specifically, when the temperature of ice-liquid coexisting system is higher than $T_m$, the chemical potential of the ice phase is higher than that of the liquid phase: $\mu_\mathrm{ice}(p,T)_{T>T_m} > \mu_\mathrm{liq}(p,T)_{T>T_m}$.
Consequently, the ice phase undergoes a process of melting with the absorption of heat, leading to a reduction in the kinetic temperature of the coexisting system towards the melting temperature $T_m$. 
One notable advantage of employing anisotropic $NpH$ MD simulations over $NVE$ MD simulation is the absence of the pressure-anisotropy problem, as the three principle components of the pressure tensor can be adjusted to the specified external pressure in the former (see Figure S5).\cite{yooMeltingLinesModel2004,yooMeltingTemperatureBulk2009,yooPhaseDiagramWater2009,vegaDeterminationPhaseDiagrams2008}

We estimate the melting temperatures of ice II, III, and V under pressure conditions of 0.2 $\sim$ 0.6 GPa using the MB-pol water model.\cite{rieraMBXManybodyEnergy2023}
To achieve this, we performed on-the-fly active learning SGPR $NpT$ MD simulations to create a BCM MLP for each coexisting system: ice II/water, ice III/water, and ice V/water. Three separately trained SGPR local expert MLP models were then combined into one sparse robust BCM potential. This combined potential was used for non-active learning $NpH$ MD simulations of coexisting ice-liquid systems. No updates were made to the inducing sets and datasets in the non-active MD simulations. Non-active $NpH$ MD simulations were performed to determine the melting temperatures of ice II, III, and V.
Throughout these MD simulations, we maintained the conservation of enthalpy ($H$), while observing fluctuations in the total energy, 
which is the sum of the potential energy and kinetic energy (Figure S6). 

Figure \ref{fig:Tm} illustrates the melting points of ice II, III, and V as determined from our $NpH$ MD simulations using the MB-pol water model. The ice-water phase diagram predicted with the MB-pol potential closely align with those obtained from enhanced-coexistence simulations, as reported by Bore and Paesani\cite{BorePaesani23} (see Figure S8).
The BCM MLP stabilizes ice V over ice II, consistent with Zhang's predicted phase diagram.\cite{zhangPhaseDiagramDeep2021} 
The Bore and Paesani\cite{BorePaesani23} using a sophisticated deep neural network potential, however, observed that ice V became a metastable phase, absent from the phase diagram. We attribute this discrepancy to the failure in incorporating high model uncertainty configurations of the water-ice system in the training dataset. In contrast, our active BCM potential adeptly learns model uncertainties on-the-fly, enhancing model accuracy and thereby stabilize ice V in the phase diagram (Figure \ref{fig:Tm}). However, the melting points predicted by the BCM MLP exhibit a slight upward shift in temperature relative to the benchmark MB-pol potential (Figure \ref{fig:Tm}). This deviation primarily stems from the noise present in the energy and force predictions of the BCM potential. Notwithstanding the shifts in melting temperatures, we emphasize the BCM potential's capability to accurately replicate the physical and thermodynamic properties of the reference MB-pol water model. While the discrepancies in melting temperatures suggest a room for further improvements in accuracy of model, the significance lies in the BCM potential's reproduction of the reference phase diagram. This affirms the BCM potential's utility in capturing the essence of water's behavior under various conditions while reducing the computational effort required to train SGPR MLP on large datasets.

Deep neural network ML models\cite{zhangPhaseDiagramDeep2021,BorePaesani23} based on DFT and MB-pol predict that ice III is metastable when compared to ice II. Our BCM MLP based on MB-pol exhibits the same trend, indicating that ice III is indeed metastable in comparison to ice II. 
The MB-pol water model, used in MD simulations, lacks the ability to allow proton transfer within ice, limiting its capacity to represent all possible configurations of proton-disordered ice. Consequently, the entropy stemming from proton disorder is not factored into the chemical potential of ice $\mu_\mathrm{ice}(p,T)$. This discrepancy extends to $NpH$ MD simulations of ice-liquid systems, where the entropy attributed to proton disorder remains unaddressed. The disparity between these models and experimental data likely arises from neglecting the entropy arising from proton disorder in ice III.\cite{macdowellCombinatorialEntropyPhase2004,vegaDeterminationPhaseDiagrams2008}
Notably, the entropy of proton-disordered ice III surpasses that of proton-ordered ice II. Correctly incorporating the entropy related to proton disorder in ice III should lead to an expectation that the melting temperature of proton-disordered ice III exceeds that of proton-ordered ice II.\cite{macdowellCombinatorialEntropyPhase2004}

Table \ref{tbl:cost} summarizes the computational costs of obtaining the weight vector $\bm{w}_m$ in Eq. (\ref{eq:wm}). 
Each SGPR local expert for ice II-water, ice III-water, and ice V-water has a comparable number of descriptors in the inducing set $z$. 
Consequently, the computational cost for acquiring $\bm{w}_m$ increases proportionally with the total number of descriptors in the dataset $\mathcal{O}(n_d)$.
Combining all SGPR local experts into a single SGPR model results in a slightly increase in both  $m$ and  $n_d$ compared to the SGPR local expert for ice II-water. 
Nevertheless, the computational cost in obtaining $\bm{w}_m$ nearly doubles in comparison with the SGPR local expert of ice II-water, satisfying that the computational cost scales as $\mathcal{O}(n_d m^2)$.

\begin{table}[h]
\small
  \caption{\ The computational cost of obtaining the solution $\bm{w}_m$ in Eq. (\ref{eq:wm}). $n$, $n_d$ and $m$ represent the number of data in the dataset, the total number of descriptors in the dataset, and the number of descriptors in the inducing set $z$. }
  \label{tbl:cost}
  \begin{tabular*}{0.48\textwidth}{@{\extracolsep{\fill}}lllll}
    \hline
    SGPR local expert & $n$ & $n_d$ & $m$ & CPU time (s) \\
    \hline
    Ice V & 13 & 26208 & 261 & 2.17 \\
    Ice III & 21 & 40824 & 300 & 4.28 \\
    Ice II & 25 & 64800 & 284 & 6.69 \\
    Ice II+Ice III+Ice V & 32 & 78624 & 357 & 11.73 \\
    \hline
  \end{tabular*}
\end{table}

\section{Conclusions}

In this study, we introduced two theoretical advancements to enhance the precision of pressure predictions and to reduce the computational scaling of SGPR MLPs. For the isobaric MD simulations (i.e. $NpT$ and $NpH$), accurate predictions of system pressure is crucial. We showed that the introduction of a virial kernel into the covariance matrix substantially improves pressure predictions, particularly for systems under extreme conditions. This improvement is important for the fidelity of isobaric MD simulations, particularly for the study of phase transition and the behavior under extreme conditions. Furthermore, the refined MLP with the virial kernel enables the estimation of the melting temperature of ice polymorphs. 
We achieve this by employing $NpH$ MD simulations of the ice-water coexisting phases for three cases (ice II, ice III, and ice V). 
Additionally, we performed active learning $NpT$ MD simulations of LGPS, demonstrating a phase transition. These results highlight the capability of our improved MLP to accurately capture phase changes in complex materials.

We also developed the active BCM potential, which addresses the computational scalability challenges inherent to $\mathcal{GP}$-based potentials, including SGPR. By aggregating local sparse SGPR kernel potentials into a unified BCM model, the active BCM potential effectively reduces the dimensionality of the kernel matrix, thus reducing the computational overhead without the need for training a large dataset. 
This approach not only streamlines the model training process but also enhances the model's transferability to a broader range of systems and conditions. Importantly, the active BCM potential facilitates on-the-fly learning of model uncertainties, further refining the model's accuracy and reliability.

These two approaches—the virial kernel and the active BCM potential—mark a significant step forward in the application of SGPR potential to isobaric MD simulations. By addressing key challenges in pressure estimation and computational efficiency, our study paves the way for more accurate, reliable, and scalable MD simulations across a wide array of materials and conditions. These advancements hold promise for deepening our understanding of material behaviors and phase transitions, contributing valuable insights to the fields of chemistry, physics, and materials science.
\newline

\section*{Author Contributions}
\noindent
\textbf{Soohaeng Yoo Willow}: Conceptualization, Methodology, Software, Validation, Formal analysis, Data Curation, Writing, Visualization, Funding acquisition. 
\noindent
\textbf{Dong Geon Kim}: Data Curation, Writing, Visualization.
\noindent
\textbf{R. Sundheep}: Data Curation, Writing.
\noindent
\textbf{Amir Hajibabaei}: Conceptualization, Software.
\noindent
\textbf{Chang Woo Myung}: Conceptualization, Methodology, Writing, Visualization, Formal analysis, Supervision, Project administration, Funding acquisition.
\newline

\section*{Conflicts of interest}

There are no conflicts to declare.

\section*{Acknowledgements}
We are grateful to V. Kapil, S. Pourasad, and C. Schran for helpful discussions.
SYW and CWM acknowledge the support from Brain Pool program funded by the Ministry of Science and ICT through the National Research Foundation of Korea (No. RS-2023-00222245).
CWM acknowledges the support from the National Research Foundation of Korea (NRF) grant funded by the Korea government (MSIT) (No. NRF-2022R1C1C1010605).
The authors are grateful for 
the computational support from the Korea Institute of Science and Technology Information (KISTI) for the Nurion cluster (KSC-2021-CRE-0542, KSC-2022-CRE-0115, KSC-2022-CRE-0424, KSC-2023-CRE-0059, KSC-2023-CRE-0198, KSC-2023-CRE-0332, KSC-2023-CRE-0355, KSC-2023-CRE-0454).
%



\section*{Data availability}
See the supplementary material for a detailed compilation of the obtained results as well as further data and analysis to support the points made throughout the text. The input and output files associated with this study and all analysis can be found on \href{https://github.com/myung-group/Data_active_BCM.git}{GitHub}. 

\balance

\bibliography{SGPR}

\providecommand*{\mcitethebibliography}{\thebibliography}
\csname @ifundefined\endcsname{endmcitethebibliography}
{\let\endmcitethebibliography\endthebibliography}{}
\begin{mcitethebibliography}{76}
\providecommand*{\natexlab}[1]{#1}
\providecommand*{\mciteSetBstSublistMode}[1]{}
\providecommand*{\mciteSetBstMaxWidthForm}[2]{}
\providecommand*{\mciteBstWouldAddEndPuncttrue}
  {\def\EndOfBibitem{\unskip.}}
\providecommand*{\mciteBstWouldAddEndPunctfalse}
  {\let\EndOfBibitem\relax}
\providecommand*{\mciteSetBstMidEndSepPunct}[3]{}
\providecommand*{\mciteSetBstSublistLabelBeginEnd}[3]{}
\providecommand*{\EndOfBibitem}{}
\mciteSetBstSublistMode{f}
\mciteSetBstMaxWidthForm{subitem}
{(\emph{\alph{mcitesubitemcount}})}
\mciteSetBstSublistLabelBeginEnd{\mcitemaxwidthsubitemform\space}
{\relax}{\relax}

\bibitem[Unke \emph{et~al.}(2021)Unke, Chmiela, Sauceda, Gastegger, Poltavsky, Sch{\"u}tt, Tkatchenko, and M{\"u}ller]{unkeMachineLearningForce2021}
O.~T. Unke, S.~Chmiela, H.~E. Sauceda, M.~Gastegger, I.~Poltavsky, K.~T. Sch{\"u}tt, A.~Tkatchenko and K.-R. M{\"u}ller, \emph{Chem. Rev.}, 2021, \textbf{121}, 10142--10186\relax
\mciteBstWouldAddEndPuncttrue
\mciteSetBstMidEndSepPunct{\mcitedefaultmidpunct}
{\mcitedefaultendpunct}{\mcitedefaultseppunct}\relax
\EndOfBibitem
\bibitem[Deringer \emph{et~al.}(2021)Deringer, Bart{\'o}k, Bernstein, Wilkins, Ceriotti, and Cs{\'a}nyi]{DeringerCsanyi21}
V.~L. Deringer, A.~P. Bart{\'o}k, N.~Bernstein, D.~M. Wilkins, M.~Ceriotti and G.~Cs{\'a}nyi, \emph{Chem. Rev.}, 2021, \textbf{121}, 10073--10141\relax
\mciteBstWouldAddEndPuncttrue
\mciteSetBstMidEndSepPunct{\mcitedefaultmidpunct}
{\mcitedefaultendpunct}{\mcitedefaultseppunct}\relax
\EndOfBibitem
\bibitem[Bart{\'o}k \emph{et~al.}(2010)Bart{\'o}k, Payne, Kondor, and Cs{\'a}nyi]{BartokCsanyi10}
A.~P. Bart{\'o}k, M.~C. Payne, R.~Kondor and G.~Cs{\'a}nyi, \emph{Phys. Rev. Lett.}, 2010, \textbf{104}, 136403\relax
\mciteBstWouldAddEndPuncttrue
\mciteSetBstMidEndSepPunct{\mcitedefaultmidpunct}
{\mcitedefaultendpunct}{\mcitedefaultseppunct}\relax
\EndOfBibitem
\bibitem[Chen \emph{et~al.}(2023)Chen, Lee, Ye, Berkelbach, Reichman, and Markland]{ChenMarkland23a}
M.~S. Chen, J.~Lee, H.-Z. Ye, T.~C. Berkelbach, D.~R. Reichman and T.~E. Markland, \emph{J. Chem. Theory Comput.}, 2023, \textbf{19}, 4510--4519\relax
\mciteBstWouldAddEndPuncttrue
\mciteSetBstMidEndSepPunct{\mcitedefaultmidpunct}
{\mcitedefaultendpunct}{\mcitedefaultseppunct}\relax
\EndOfBibitem
\bibitem[Gilmer \emph{et~al.}(2017)Gilmer, Schoenholz, Riley, Vinyals, and Dahl]{gilmerNeuralMessagePassing2017}
J.~Gilmer, S.~S. Schoenholz, P.~F. Riley, O.~Vinyals and G.~E. Dahl, \emph{Neural {{Message Passing}} for {{Quantum Chemistry}}}, 2017\relax
\mciteBstWouldAddEndPuncttrue
\mciteSetBstMidEndSepPunct{\mcitedefaultmidpunct}
{\mcitedefaultendpunct}{\mcitedefaultseppunct}\relax
\EndOfBibitem
\bibitem[Batzner \emph{et~al.}(2022)Batzner, Musaelian, Sun, Geiger, Mailoa, Kornbluth, Molinari, Smidt, and Kozinsky]{batznerEquivariantGraphNeural2022}
S.~Batzner, A.~Musaelian, L.~Sun, M.~Geiger, J.~P. Mailoa, M.~Kornbluth, N.~Molinari, T.~E. Smidt and B.~Kozinsky, \emph{Nat. Commun.}, 2022, \textbf{13}, 2453\relax
\mciteBstWouldAddEndPuncttrue
\mciteSetBstMidEndSepPunct{\mcitedefaultmidpunct}
{\mcitedefaultendpunct}{\mcitedefaultseppunct}\relax
\EndOfBibitem
\bibitem[Hajibabaei \emph{et~al.}(2021)Hajibabaei, Myung, and Kim]{hajibabaeiSparseGaussianProcess2021}
A.~Hajibabaei, C.~W. Myung and K.~S. Kim, \emph{Phys. Rev. B}, 2021, \textbf{103}, 214102\relax
\mciteBstWouldAddEndPuncttrue
\mciteSetBstMidEndSepPunct{\mcitedefaultmidpunct}
{\mcitedefaultendpunct}{\mcitedefaultseppunct}\relax
\EndOfBibitem
\bibitem[Hajibabaei and Kim(2021)]{hajibabaeiUniversalMachineLearning2021}
A.~Hajibabaei and K.~S. Kim, \emph{J. Phys. Chem. Lett.}, 2021, \textbf{12}, 8115--8120\relax
\mciteBstWouldAddEndPuncttrue
\mciteSetBstMidEndSepPunct{\mcitedefaultmidpunct}
{\mcitedefaultendpunct}{\mcitedefaultseppunct}\relax
\EndOfBibitem
\bibitem[Hajibabaei \emph{et~al.}(2021)Hajibabaei, Ha, Pourasad, Kim, and Kim]{HajibabaeiKim21b}
A.~Hajibabaei, M.~Ha, S.~Pourasad, J.~Kim and K.~S. Kim, \emph{J. Phys. Chem. A}, 2021, \textbf{125}, 9414--9420\relax
\mciteBstWouldAddEndPuncttrue
\mciteSetBstMidEndSepPunct{\mcitedefaultmidpunct}
{\mcitedefaultendpunct}{\mcitedefaultseppunct}\relax
\EndOfBibitem
\bibitem[Myung \emph{et~al.}(2022)Myung, Hajibabaei, Cha, Ha, Kim, and Kim]{myungChallengesOpportunitiesProspects2022}
C.~W. Myung, A.~Hajibabaei, J.-H. Cha, M.~Ha, J.~Kim and K.~S. Kim, \emph{Adv. Energy Mater.}, 2022, \textbf{12}, 2202279\relax
\mciteBstWouldAddEndPuncttrue
\mciteSetBstMidEndSepPunct{\mcitedefaultmidpunct}
{\mcitedefaultendpunct}{\mcitedefaultseppunct}\relax
\EndOfBibitem
\bibitem[Vandermause \emph{et~al.}(2022)Vandermause, Xie, Lim, Owen, and Kozinsky]{VandermauseKozinsky22}
J.~Vandermause, Y.~Xie, J.~S. Lim, C.~J. Owen and B.~Kozinsky, \emph{Nat. Commun.}, 2022, \textbf{13}, 5183\relax
\mciteBstWouldAddEndPuncttrue
\mciteSetBstMidEndSepPunct{\mcitedefaultmidpunct}
{\mcitedefaultendpunct}{\mcitedefaultseppunct}\relax
\EndOfBibitem
\bibitem[Vandermause \emph{et~al.}(2020)Vandermause, Torrisi, Batzner, Xie, Sun, Kolpak, and Kozinsky]{VandermauseKozinsky20}
J.~Vandermause, S.~B. Torrisi, S.~Batzner, Y.~Xie, L.~Sun, A.~M. Kolpak and B.~Kozinsky, \emph{npj Comput. Mater.}, 2020, \textbf{6}, 20\relax
\mciteBstWouldAddEndPuncttrue
\mciteSetBstMidEndSepPunct{\mcitedefaultmidpunct}
{\mcitedefaultendpunct}{\mcitedefaultseppunct}\relax
\EndOfBibitem
\bibitem[Metcalf \emph{et~al.}(2020)Metcalf, Koutsoukas, Spronk, Claus, Loughney, Johnson, Cheney, and Sherrill]{metcalfApproachesMachineLearning2020}
D.~P. Metcalf, A.~Koutsoukas, S.~A. Spronk, B.~L. Claus, D.~A. Loughney, S.~R. Johnson, D.~L. Cheney and C.~D. Sherrill, \emph{J. Chem. Phys.}, 2020, \textbf{152}, 074103\relax
\mciteBstWouldAddEndPuncttrue
\mciteSetBstMidEndSepPunct{\mcitedefaultmidpunct}
{\mcitedefaultendpunct}{\mcitedefaultseppunct}\relax
\EndOfBibitem
\bibitem[Na and Kim(2021)]{naReverseGraphSelfattention2021}
G.~S. Na and H.~W. Kim, \emph{Neural Netw.}, 2021, \textbf{133}, 1--10\relax
\mciteBstWouldAddEndPuncttrue
\mciteSetBstMidEndSepPunct{\mcitedefaultmidpunct}
{\mcitedefaultendpunct}{\mcitedefaultseppunct}\relax
\EndOfBibitem
\bibitem[Zhung \emph{et~al.}(2024)Zhung, Kim, and Kim]{zhung3DMolecularGenerative2024}
W.~Zhung, H.~Kim and W.~Y. Kim, \emph{Nat Commun}, 2024, \textbf{15}, 2688\relax
\mciteBstWouldAddEndPuncttrue
\mciteSetBstMidEndSepPunct{\mcitedefaultmidpunct}
{\mcitedefaultendpunct}{\mcitedefaultseppunct}\relax
\EndOfBibitem
\bibitem[Imbalzano \emph{et~al.}(2021)Imbalzano, Zhuang, Kapil, Rossi, Engel, Grasselli, and Ceriotti]{imbalzanoUncertaintyEstimationMolecular2021}
G.~Imbalzano, Y.~Zhuang, V.~Kapil, K.~Rossi, E.~A. Engel, F.~Grasselli and M.~Ceriotti, \emph{J. Chem. Phys.}, 2021, \textbf{154}, 074102\relax
\mciteBstWouldAddEndPuncttrue
\mciteSetBstMidEndSepPunct{\mcitedefaultmidpunct}
{\mcitedefaultendpunct}{\mcitedefaultseppunct}\relax
\EndOfBibitem
\bibitem[Bayerl \emph{et~al.}(2022)Bayerl, Andolina, Dwaraknath, and Saidi]{Bayerl2022}
D.~Bayerl, C.~M. Andolina, S.~Dwaraknath and W.~A. Saidi, \emph{Digital Discovery}, 2022, \textbf{1}, 61--69\relax
\mciteBstWouldAddEndPuncttrue
\mciteSetBstMidEndSepPunct{\mcitedefaultmidpunct}
{\mcitedefaultendpunct}{\mcitedefaultseppunct}\relax
\EndOfBibitem
\bibitem[Laasonen \emph{et~al.}(1993)Laasonen, Sprik, Parrinello, and Car]{laasonenInitioLiquidWater1993}
K.~Laasonen, M.~Sprik, M.~Parrinello and R.~Car, \emph{J. Chem. Phys.}, 1993, \textbf{99}, 9080--9089\relax
\mciteBstWouldAddEndPuncttrue
\mciteSetBstMidEndSepPunct{\mcitedefaultmidpunct}
{\mcitedefaultendpunct}{\mcitedefaultseppunct}\relax
\EndOfBibitem
\bibitem[Richards \emph{et~al.}(2016)Richards, Tsujimura, Miara, Wang, Kim, Ong, Uechi, Suzuki, and Ceder]{richardsDesignSynthesisSuperionic2016}
W.~D. Richards, T.~Tsujimura, L.~J. Miara, Y.~Wang, J.~C. Kim, S.~P. Ong, I.~Uechi, N.~Suzuki and G.~Ceder, \emph{Nat. Commun.}, 2016, \textbf{7}, 11009\relax
\mciteBstWouldAddEndPuncttrue
\mciteSetBstMidEndSepPunct{\mcitedefaultmidpunct}
{\mcitedefaultendpunct}{\mcitedefaultseppunct}\relax
\EndOfBibitem
\bibitem[Boero \emph{et~al.}(1998)Boero, Parrinello, and Terakura]{boeroFirstPrinciplesMolecular1998}
M.~Boero, M.~Parrinello and K.~Terakura, \emph{J. Am. Chem. Soc.}, 1998, \textbf{120}, 2746--2752\relax
\mciteBstWouldAddEndPuncttrue
\mciteSetBstMidEndSepPunct{\mcitedefaultmidpunct}
{\mcitedefaultendpunct}{\mcitedefaultseppunct}\relax
\EndOfBibitem
\bibitem[Willow \emph{et~al.}(2015)Willow, Salim, Kim, and Hirata]{WillowHirata15}
S.~Y. Willow, M.~A. Salim, K.~S. Kim and S.~Hirata, \emph{Sci. Rep.}, 2015, \textbf{5}, 14358\relax
\mciteBstWouldAddEndPuncttrue
\mciteSetBstMidEndSepPunct{\mcitedefaultmidpunct}
{\mcitedefaultendpunct}{\mcitedefaultseppunct}\relax
\EndOfBibitem
\bibitem[Del~Ben \emph{et~al.}(2013)Del~Ben, Sch{\"o}nherr, Hutter, and VandeVondele]{delbenBulkLiquidWater2013}
M.~Del~Ben, M.~Sch{\"o}nherr, J.~Hutter and J.~VandeVondele, \emph{J. Phys. Chem. Lett.}, 2013, \textbf{4}, 3753--3759\relax
\mciteBstWouldAddEndPuncttrue
\mciteSetBstMidEndSepPunct{\mcitedefaultmidpunct}
{\mcitedefaultendpunct}{\mcitedefaultseppunct}\relax
\EndOfBibitem
\bibitem[Del~Ben \emph{et~al.}(2015)Del~Ben, Hutter, and VandeVondele]{delbenForcesStressSecond2015}
M.~Del~Ben, J.~Hutter and J.~VandeVondele, \emph{J. Chem. Phys.}, 2015, \textbf{143}, 102803\relax
\mciteBstWouldAddEndPuncttrue
\mciteSetBstMidEndSepPunct{\mcitedefaultmidpunct}
{\mcitedefaultendpunct}{\mcitedefaultseppunct}\relax
\EndOfBibitem
\bibitem[Behler and Parrinello(2007)]{behlerGeneralizedNeuralNetworkRepresentation2007}
J.~Behler and M.~Parrinello, \emph{Phys. Rev. Lett.}, 2007, \textbf{98}, 146401\relax
\mciteBstWouldAddEndPuncttrue
\mciteSetBstMidEndSepPunct{\mcitedefaultmidpunct}
{\mcitedefaultendpunct}{\mcitedefaultseppunct}\relax
\EndOfBibitem
\bibitem[Behler(2011)]{behlerAtomcenteredSymmetryFunctions2011}
J.~Behler, \emph{J. Chem. Phys.}, 2011, \textbf{134}, 074106\relax
\mciteBstWouldAddEndPuncttrue
\mciteSetBstMidEndSepPunct{\mcitedefaultmidpunct}
{\mcitedefaultendpunct}{\mcitedefaultseppunct}\relax
\EndOfBibitem
\bibitem[Pun \emph{et~al.}(2019)Pun, Batra, Ramprasad, and Mishin]{punPhysicallyInformedArtificial2019}
G.~P.~P. Pun, R.~Batra, R.~Ramprasad and Y.~Mishin, \emph{Nat. Commun.}, 2019, \textbf{10}, 2339\relax
\mciteBstWouldAddEndPuncttrue
\mciteSetBstMidEndSepPunct{\mcitedefaultmidpunct}
{\mcitedefaultendpunct}{\mcitedefaultseppunct}\relax
\EndOfBibitem
\bibitem[Eckhoff and Behler(2021)]{eckhoffHighdimensionalNeuralNetwork2021}
M.~Eckhoff and J.~Behler, \emph{Npj Comput. Mater.}, 2021, \textbf{7}, 170\relax
\mciteBstWouldAddEndPuncttrue
\mciteSetBstMidEndSepPunct{\mcitedefaultmidpunct}
{\mcitedefaultendpunct}{\mcitedefaultseppunct}\relax
\EndOfBibitem
\bibitem[Schran \emph{et~al.}(2020)Schran, Brezina, and Marsalek]{schranCommitteeNeuralNetwork2020}
C.~Schran, K.~Brezina and O.~Marsalek, \emph{J. Chem. Phys.}, 2020, \textbf{153}, 104105\relax
\mciteBstWouldAddEndPuncttrue
\mciteSetBstMidEndSepPunct{\mcitedefaultmidpunct}
{\mcitedefaultendpunct}{\mcitedefaultseppunct}\relax
\EndOfBibitem
\bibitem[Schran \emph{et~al.}(2021)Schran, Thiemann, Rowe, M{\"u}ller, Marsalek, and Michaelides]{schranMachineLearningPotentials2021}
C.~Schran, F.~L. Thiemann, P.~Rowe, E.~A. M{\"u}ller, O.~Marsalek and A.~Michaelides, \emph{Proc. Natl. Acad. Sci. U.S.A.}, 2021, \textbf{118}, e2110077118\relax
\mciteBstWouldAddEndPuncttrue
\mciteSetBstMidEndSepPunct{\mcitedefaultmidpunct}
{\mcitedefaultendpunct}{\mcitedefaultseppunct}\relax
\EndOfBibitem
\bibitem[Kapil \emph{et~al.}(2022)Kapil, Schran, Zen, Chen, Pickard, and Michaelides]{kapilFirstprinciplesPhaseDiagram2022}
V.~Kapil, C.~Schran, A.~Zen, J.~Chen, C.~J. Pickard and A.~Michaelides, \emph{Nature}, 2022, \textbf{609}, 512--516\relax
\mciteBstWouldAddEndPuncttrue
\mciteSetBstMidEndSepPunct{\mcitedefaultmidpunct}
{\mcitedefaultendpunct}{\mcitedefaultseppunct}\relax
\EndOfBibitem
\bibitem[Bart{\'o}k \emph{et~al.}(2013)Bart{\'o}k, Kondor, and Cs{\'a}nyi]{bartokRepresentingChemicalEnvironments2013}
A.~P. Bart{\'o}k, R.~Kondor and G.~Cs{\'a}nyi, \emph{Phys. Rev. B}, 2013, \textbf{87}, 184115\relax
\mciteBstWouldAddEndPuncttrue
\mciteSetBstMidEndSepPunct{\mcitedefaultmidpunct}
{\mcitedefaultendpunct}{\mcitedefaultseppunct}\relax
\EndOfBibitem
\bibitem[Bart{\'o}k and Cs{\'a}nyi(2015)]{bartokAussianApproximationPotentials2015}
A.~P. Bart{\'o}k and G.~Cs{\'a}nyi, \emph{Int. J. of Quantum Chemistry}, 2015, \textbf{115}, 1051--1057\relax
\mciteBstWouldAddEndPuncttrue
\mciteSetBstMidEndSepPunct{\mcitedefaultmidpunct}
{\mcitedefaultendpunct}{\mcitedefaultseppunct}\relax
\EndOfBibitem
\bibitem[Klawohn \emph{et~al.}(2023)Klawohn, Darby, Kermode, Cs{\'a}nyi, Caro, and Bart{\'o}k]{klawohnGaussianApproximationPotentials2023}
S.~Klawohn, J.~P. Darby, J.~R. Kermode, G.~Cs{\'a}nyi, M.~A. Caro and A.~P. Bart{\'o}k, \emph{J. Chem. Phys.}, 2023, \textbf{159}, 174108\relax
\mciteBstWouldAddEndPuncttrue
\mciteSetBstMidEndSepPunct{\mcitedefaultmidpunct}
{\mcitedefaultendpunct}{\mcitedefaultseppunct}\relax
\EndOfBibitem
\bibitem[{Qui{\~n}onero-Candela} and Rasmussen(2005)]{quinonero-candelaUnifyingViewSparse2005}
J.~{Qui{\~n}onero-Candela} and C.~E. Rasmussen, \emph{J. Mach. Learn. Res.}, 2005, \textbf{6}, 1939--1959\relax
\mciteBstWouldAddEndPuncttrue
\mciteSetBstMidEndSepPunct{\mcitedefaultmidpunct}
{\mcitedefaultendpunct}{\mcitedefaultseppunct}\relax
\EndOfBibitem
\bibitem[Chmiela \emph{et~al.}(2017)Chmiela, Tkatchenko, Sauceda, Poltavsky, Sch{\"u}tt, and M{\"u}ller]{chmielaMachineLearningAccurate2017}
S.~Chmiela, A.~Tkatchenko, H.~E. Sauceda, I.~Poltavsky, K.~T. Sch{\"u}tt and K.-R. M{\"u}ller, \emph{Sci. Adv.}, 2017, \textbf{3}, e1603015\relax
\mciteBstWouldAddEndPuncttrue
\mciteSetBstMidEndSepPunct{\mcitedefaultmidpunct}
{\mcitedefaultendpunct}{\mcitedefaultseppunct}\relax
\EndOfBibitem
\bibitem[Shapeev(2016)]{shapeevMomentTensorPotentials2016}
A.~V. Shapeev, \emph{Multiscale Model. Simul.}, 2016, \textbf{14}, 1153--1173\relax
\mciteBstWouldAddEndPuncttrue
\mciteSetBstMidEndSepPunct{\mcitedefaultmidpunct}
{\mcitedefaultendpunct}{\mcitedefaultseppunct}\relax
\EndOfBibitem
\bibitem[Podryabinkin and Shapeev(2017)]{podryabinkinActiveLearningLinearly2017}
E.~V. Podryabinkin and A.~V. Shapeev, \emph{Comput. Mater. Sci.}, 2017, \textbf{140}, 171--180\relax
\mciteBstWouldAddEndPuncttrue
\mciteSetBstMidEndSepPunct{\mcitedefaultmidpunct}
{\mcitedefaultendpunct}{\mcitedefaultseppunct}\relax
\EndOfBibitem
\bibitem[Novikov \emph{et~al.}(2021)Novikov, Gubaev, Podryabinkin, and Shapeev]{NovikovShapeev21}
I.~S. Novikov, K.~Gubaev, E.~V. Podryabinkin and A.~V. Shapeev, \emph{Mach. Learn.: Sci. Technol.}, 2021, \textbf{2}, 025002\relax
\mciteBstWouldAddEndPuncttrue
\mciteSetBstMidEndSepPunct{\mcitedefaultmidpunct}
{\mcitedefaultendpunct}{\mcitedefaultseppunct}\relax
\EndOfBibitem
\bibitem[Batatia \emph{et~al.}(2022)Batatia, Kovacs, Simm, Ortner, and Csanyi]{NEURIPS2022_4a36c3c5}
I.~Batatia, D.~P. Kovacs, G.~Simm, C.~Ortner and G.~Csanyi, \emph{Advances in Neural Information Processing Systems}, 2022, \textbf{35}, 11423--11436\relax
\mciteBstWouldAddEndPuncttrue
\mciteSetBstMidEndSepPunct{\mcitedefaultmidpunct}
{\mcitedefaultendpunct}{\mcitedefaultseppunct}\relax
\EndOfBibitem
\bibitem[Lim and Jung(2021)]{lim_mlsolva_2021}
H.~Lim and Y.~Jung, \emph{J. Cheminform.}, 2021, \textbf{13}, 56\relax
\mciteBstWouldAddEndPuncttrue
\mciteSetBstMidEndSepPunct{\mcitedefaultmidpunct}
{\mcitedefaultendpunct}{\mcitedefaultseppunct}\relax
\EndOfBibitem
\bibitem[Rasmussen and Williams(2005)]{rasmussenGaussianProcessesMachine2005}
C.~E. Rasmussen and C.~K.~I. Williams, \emph{Gaussian {{Processes}} for {{Machine Learning}}}, The MIT Press, 2005\relax
\mciteBstWouldAddEndPuncttrue
\mciteSetBstMidEndSepPunct{\mcitedefaultmidpunct}
{\mcitedefaultendpunct}{\mcitedefaultseppunct}\relax
\EndOfBibitem
\bibitem[Shahriari \emph{et~al.}(2016)Shahriari, Swersky, Wang, Adams, and De~Freitas]{shahriariTakingHumanOut2016}
B.~Shahriari, K.~Swersky, Z.~Wang, R.~P. Adams and N.~De~Freitas, \emph{Proc. IEEE}, 2016, \textbf{104}, 148--175\relax
\mciteBstWouldAddEndPuncttrue
\mciteSetBstMidEndSepPunct{\mcitedefaultmidpunct}
{\mcitedefaultendpunct}{\mcitedefaultseppunct}\relax
\EndOfBibitem
\bibitem[Lawrence(2005)]{lawrenceProbabilisticNonlinearPrincipal2005}
N.~Lawrence, \emph{J. Mach. Learn. Res.}, 2005, \textbf{6}, 1783--1816\relax
\mciteBstWouldAddEndPuncttrue
\mciteSetBstMidEndSepPunct{\mcitedefaultmidpunct}
{\mcitedefaultendpunct}{\mcitedefaultseppunct}\relax
\EndOfBibitem
\bibitem[{\'A}lvarez and Lawrence(2011)]{alvarezComputationallyEfficientConvolved2011}
M.~A. {\'A}lvarez and N.~D. Lawrence, \emph{J. Mach. Learn. Res.}, 2011, \textbf{12}, 1459--1500\relax
\mciteBstWouldAddEndPuncttrue
\mciteSetBstMidEndSepPunct{\mcitedefaultmidpunct}
{\mcitedefaultendpunct}{\mcitedefaultseppunct}\relax
\EndOfBibitem
\bibitem[Tresp(2000)]{trespBayesianCommitteeMachine2000}
V.~Tresp, \emph{Neural Computation}, 2000, \textbf{12}, 2719--2741\relax
\mciteBstWouldAddEndPuncttrue
\mciteSetBstMidEndSepPunct{\mcitedefaultmidpunct}
{\mcitedefaultendpunct}{\mcitedefaultseppunct}\relax
\EndOfBibitem
\bibitem[Wang \emph{et~al.}(2018)Wang, Zhang, Han, and E]{wangDeePMDkitDeepLearning2018}
H.~Wang, L.~Zhang, J.~Han and W.~E, \emph{Comput. Phys. Commun.}, 2018, \textbf{228}, 178--184\relax
\mciteBstWouldAddEndPuncttrue
\mciteSetBstMidEndSepPunct{\mcitedefaultmidpunct}
{\mcitedefaultendpunct}{\mcitedefaultseppunct}\relax
\EndOfBibitem
\bibitem[Zhang \emph{et~al.}(2021)Zhang, Wang, Car, and E]{zhangPhaseDiagramDeep2021}
L.~Zhang, H.~Wang, R.~Car and W.~E, \emph{Phys. Rev. Lett.}, 2021, \textbf{126}, 236001\relax
\mciteBstWouldAddEndPuncttrue
\mciteSetBstMidEndSepPunct{\mcitedefaultmidpunct}
{\mcitedefaultendpunct}{\mcitedefaultseppunct}\relax
\EndOfBibitem
\bibitem[Sun \emph{et~al.}(2015)Sun, Ruzsinszky, and Perdew]{sunStronglyConstrainedAppropriately2015}
J.~Sun, A.~Ruzsinszky and J.~P. Perdew, \emph{Phys. Rev. Lett.}, 2015, \textbf{115}, 036402\relax
\mciteBstWouldAddEndPuncttrue
\mciteSetBstMidEndSepPunct{\mcitedefaultmidpunct}
{\mcitedefaultendpunct}{\mcitedefaultseppunct}\relax
\EndOfBibitem
\bibitem[Zhang \emph{et~al.}(2019)Zhang, Lin, Wang, Car, and E]{zhangActiveLearningUniformly2019}
L.~Zhang, D.-Y. Lin, H.~Wang, R.~Car and W.~E, \emph{Phys. Rev. Mater.}, 2019, \textbf{3}, 023804\relax
\mciteBstWouldAddEndPuncttrue
\mciteSetBstMidEndSepPunct{\mcitedefaultmidpunct}
{\mcitedefaultendpunct}{\mcitedefaultseppunct}\relax
\EndOfBibitem
\bibitem[Bore and Paesani(2023)]{BorePaesani23}
S.~L. Bore and F.~Paesani, \emph{Nat. Commun.}, 2023, \textbf{14}, 3349\relax
\mciteBstWouldAddEndPuncttrue
\mciteSetBstMidEndSepPunct{\mcitedefaultmidpunct}
{\mcitedefaultendpunct}{\mcitedefaultseppunct}\relax
\EndOfBibitem
\bibitem[Babin \emph{et~al.}(2013)Babin, Leforestier, and Paesani]{babinDevelopmentFirstPrinciples2013}
V.~Babin, C.~Leforestier and F.~Paesani, \emph{J. Chem. Theory Comput.}, 2013, \textbf{9}, 5395--5403\relax
\mciteBstWouldAddEndPuncttrue
\mciteSetBstMidEndSepPunct{\mcitedefaultmidpunct}
{\mcitedefaultendpunct}{\mcitedefaultseppunct}\relax
\EndOfBibitem
\bibitem[Babin \emph{et~al.}(2014)Babin, Medders, and Paesani]{babinDevelopmentFirstPrinciples2014}
V.~Babin, G.~R. Medders and F.~Paesani, \emph{J. Chem. Theory Comput.}, 2014, \textbf{10}, 1599--1607\relax
\mciteBstWouldAddEndPuncttrue
\mciteSetBstMidEndSepPunct{\mcitedefaultmidpunct}
{\mcitedefaultendpunct}{\mcitedefaultseppunct}\relax
\EndOfBibitem
\bibitem[Riera \emph{et~al.}(2023)Riera, Knight, {Bull-Vulpe}, Zhu, Agnew, Smith, Simmonett, and Paesani]{rieraMBXManybodyEnergy2023}
M.~Riera, C.~Knight, E.~F. {Bull-Vulpe}, X.~Zhu, H.~Agnew, D.~G.~A. Smith, A.~C. Simmonett and F.~Paesani, \emph{J. Chem. Phys.}, 2023, \textbf{159}, 054802\relax
\mciteBstWouldAddEndPuncttrue
\mciteSetBstMidEndSepPunct{\mcitedefaultmidpunct}
{\mcitedefaultendpunct}{\mcitedefaultseppunct}\relax
\EndOfBibitem
\bibitem[Larsen \emph{et~al.}(2017)Larsen, Mortensen, Blomqvist, Castelli, Christensen, Dułak, Friis, Groves, Hammer, Hargus, Hermes, Jennings, Jensen, Kermode, Kitchin, Kolsbjerg, Kubal, Kaasbjerg, Lysgaard, Maronsson, Maxson, Olsen, Pastewka, Peterson, Rostgaard, Schiøtz, Schütt, Strange, Thygesen, Vegge, Vilhelmsen, Walter, Zeng, and Jacobsen]{larsen_atomic_2017}
A.~H. Larsen, J.~J. Mortensen, J.~Blomqvist, I.~E. Castelli, R.~Christensen, M.~Dułak, J.~Friis, M.~N. Groves, B.~Hammer, C.~Hargus, E.~D. Hermes, P.~C. Jennings, P.~B. Jensen, J.~Kermode, J.~R. Kitchin, E.~L. Kolsbjerg, J.~Kubal, K.~Kaasbjerg, S.~Lysgaard, J.~B. Maronsson, T.~Maxson, T.~Olsen, L.~Pastewka, A.~Peterson, C.~Rostgaard, J.~Schiøtz, O.~Schütt, M.~Strange, K.~S. Thygesen, T.~Vegge, L.~Vilhelmsen, M.~Walter, Z.~Zeng and K.~W. Jacobsen, \emph{J. Phys. Condens. Matter}, 2017, \textbf{29}, 273002\relax
\mciteBstWouldAddEndPuncttrue
\mciteSetBstMidEndSepPunct{\mcitedefaultmidpunct}
{\mcitedefaultendpunct}{\mcitedefaultseppunct}\relax
\EndOfBibitem
\bibitem[Melchionna \emph{et~al.}(1993)Melchionna, Ciccotti, and Holian]{melchionnaHooverNPTDynamics1993}
S.~Melchionna, G.~Ciccotti and B.~L. Holian, \emph{Mol. Phys.}, 1993, \textbf{78}, 533--544\relax
\mciteBstWouldAddEndPuncttrue
\mciteSetBstMidEndSepPunct{\mcitedefaultmidpunct}
{\mcitedefaultendpunct}{\mcitedefaultseppunct}\relax
\EndOfBibitem
\bibitem[Melchionna(2000)]{melchionnaConstrainedSystemsStatistical2000}
S.~Melchionna, \emph{Phys. Rev. E}, 2000, \textbf{61}, 6165--6170\relax
\mciteBstWouldAddEndPuncttrue
\mciteSetBstMidEndSepPunct{\mcitedefaultmidpunct}
{\mcitedefaultendpunct}{\mcitedefaultseppunct}\relax
\EndOfBibitem
\bibitem[Holian \emph{et~al.}(1990)Holian, De~Groot, Hoover, and Hoover]{holianTimereversibleEquilibriumNonequilibrium1990}
B.~L. Holian, A.~J. De~Groot, W.~G. Hoover and C.~G. Hoover, \emph{Phys. Rev. A}, 1990, \textbf{41}, 4552--4553\relax
\mciteBstWouldAddEndPuncttrue
\mciteSetBstMidEndSepPunct{\mcitedefaultmidpunct}
{\mcitedefaultendpunct}{\mcitedefaultseppunct}\relax
\EndOfBibitem
\bibitem[Furness \emph{et~al.}(2020)Furness, Kaplan, Ning, Perdew, and Sun]{r2scan}
J.~W. Furness, A.~D. Kaplan, J.~Ning, J.~P. Perdew and J.~Sun, \emph{J. Phys. Chem. Lett.}, 2020, \textbf{11}, 8208--8215\relax
\mciteBstWouldAddEndPuncttrue
\mciteSetBstMidEndSepPunct{\mcitedefaultmidpunct}
{\mcitedefaultendpunct}{\mcitedefaultseppunct}\relax
\EndOfBibitem
\bibitem[Caldeweyher \emph{et~al.}(2017)Caldeweyher, Bannwarth, and Grimme]{dftd4}
E.~Caldeweyher, C.~Bannwarth and S.~Grimme, \emph{J. Chem. Phys.}, 2017, \textbf{147}, 034112\relax
\mciteBstWouldAddEndPuncttrue
\mciteSetBstMidEndSepPunct{\mcitedefaultmidpunct}
{\mcitedefaultendpunct}{\mcitedefaultseppunct}\relax
\EndOfBibitem
\bibitem[Perdew \emph{et~al.}(1996)Perdew, Ernzerhof, and Burke]{perdewRationaleMixingExact1996}
J.~P. Perdew, M.~Ernzerhof and K.~Burke, \emph{J. Chem. Phys.}, 1996, \textbf{105}, 9982--9985\relax
\mciteBstWouldAddEndPuncttrue
\mciteSetBstMidEndSepPunct{\mcitedefaultmidpunct}
{\mcitedefaultendpunct}{\mcitedefaultseppunct}\relax
\EndOfBibitem
\bibitem[Kresse and Joubert(1999)]{PAW}
G.~Kresse and D.~Joubert, \emph{Phys. Rev. B}, 1999, \textbf{59}, 1758--1775\relax
\mciteBstWouldAddEndPuncttrue
\mciteSetBstMidEndSepPunct{\mcitedefaultmidpunct}
{\mcitedefaultendpunct}{\mcitedefaultseppunct}\relax
\EndOfBibitem
\bibitem[Kresse and Furthm{\"u}ller(1996)]{PhysRevB.54.11169}
G.~Kresse and J.~Furthm{\"u}ller, \emph{Phys. Rev. B}, 1996, \textbf{54}, 11169--11186\relax
\mciteBstWouldAddEndPuncttrue
\mciteSetBstMidEndSepPunct{\mcitedefaultmidpunct}
{\mcitedefaultendpunct}{\mcitedefaultseppunct}\relax
\EndOfBibitem
\bibitem[Kresse and Furthm{\"u}ller(1996)]{kresseEfficiencyAbinitioTotal1996}
G.~Kresse and J.~Furthm{\"u}ller, \emph{Comput. Mater. Sci.}, 1996, \textbf{6}, 15--50\relax
\mciteBstWouldAddEndPuncttrue
\mciteSetBstMidEndSepPunct{\mcitedefaultmidpunct}
{\mcitedefaultendpunct}{\mcitedefaultseppunct}\relax
\EndOfBibitem
\bibitem[Zhang \emph{et~al.}(2018)Zhang, Han, Wang, Saidi, Car, and E]{zhang2018nips}
L.~Zhang, J.~Han, H.~Wang, W.~Saidi, R.~Car and W.~E, \emph{Advances in Neural Information Processing Systems}, 2018, \textbf{31}, \relax
\mciteBstWouldAddEndPuncttrue
\mciteSetBstMidEndSepPunct{\mcitedefaultmidpunct}
{\mcitedefaultendpunct}{\mcitedefaultseppunct}\relax
\EndOfBibitem
\bibitem[Hong \emph{et~al.}(2020)Hong, Lee, Lee, Lee, Ma, Kim, Yoon, Ihm, Kim, Shin, Kim, Jeon, Jeon, Kim, Lee, Lee, Antidormi, Roche, Chhowalla, Shin, and Shin]{hong_ultralow-dielectric-constant_2020}
S.~Hong, C.-S. Lee, M.-H. Lee, Y.~Lee, K.~Y. Ma, G.~Kim, S.~I. Yoon, K.~Ihm, K.-J. Kim, T.~J. Shin, S.~W. Kim, E.-c. Jeon, H.~Jeon, J.-Y. Kim, H.-I. Lee, Z.~Lee, A.~Antidormi, S.~Roche, M.~Chhowalla, H.-J. Shin and H.~S. Shin, \emph{Nature}, 2020, \textbf{582}, 511--514\relax
\mciteBstWouldAddEndPuncttrue
\mciteSetBstMidEndSepPunct{\mcitedefaultmidpunct}
{\mcitedefaultendpunct}{\mcitedefaultseppunct}\relax
\EndOfBibitem
\bibitem[Mo \emph{et~al.}(2012)Mo, Ong, and Ceder]{moFirstPrinciplesStudy2012}
Y.~Mo, S.~P. Ong and G.~Ceder, \emph{Chem. Mater.}, 2012, \textbf{24}, 15--17\relax
\mciteBstWouldAddEndPuncttrue
\mciteSetBstMidEndSepPunct{\mcitedefaultmidpunct}
{\mcitedefaultendpunct}{\mcitedefaultseppunct}\relax
\EndOfBibitem
\bibitem[Hori \emph{et~al.}(2015)Hori, Kato, Suzuki, Hirayama, Kato, and Kanno]{horiPhaseDiagramLGPS2015}
S.~Hori, M.~Kato, K.~Suzuki, M.~Hirayama, Y.~Kato and R.~Kanno, \emph{J. Am. Ceram. Soc.}, 2015, \textbf{98}, 3352--3360\relax
\mciteBstWouldAddEndPuncttrue
\mciteSetBstMidEndSepPunct{\mcitedefaultmidpunct}
{\mcitedefaultendpunct}{\mcitedefaultseppunct}\relax
\EndOfBibitem
\bibitem[Wisesa \emph{et~al.}(2023)Wisesa, Andolina, and Saidi]{Wisesa2023}
P.~Wisesa, C.~M. Andolina and W.~A. Saidi, \emph{J. Phys. Chem. Lett.}, 2023, \textbf{14}, 468--475\relax
\mciteBstWouldAddEndPuncttrue
\mciteSetBstMidEndSepPunct{\mcitedefaultmidpunct}
{\mcitedefaultendpunct}{\mcitedefaultseppunct}\relax
\EndOfBibitem
\bibitem[Wisesa \emph{et~al.}(2023)Wisesa, Andolina, and Saidi]{Wisesa2023-2}
P.~Wisesa, C.~M. Andolina and W.~A. Saidi, \emph{J. Phys. Chem. Lett.}, 2023, \textbf{14}, 8741--8748\relax
\mciteBstWouldAddEndPuncttrue
\mciteSetBstMidEndSepPunct{\mcitedefaultmidpunct}
{\mcitedefaultendpunct}{\mcitedefaultseppunct}\relax
\EndOfBibitem
\bibitem[Andolina and Saidi(2023)]{andolinaTransferableMLP2023}
C.~M. Andolina and W.~A. Saidi, \emph{Digital Discovery}, 2023, \textbf{2}, 1070--1077\relax
\mciteBstWouldAddEndPuncttrue
\mciteSetBstMidEndSepPunct{\mcitedefaultmidpunct}
{\mcitedefaultendpunct}{\mcitedefaultseppunct}\relax
\EndOfBibitem
\bibitem[Petrenko and Whitworth(2002)]{Ice_Exp}
V.~F. Petrenko and R.~W. Whitworth, \emph{{Physics of Ice}}, Oxford University Press, 2002\relax
\mciteBstWouldAddEndPuncttrue
\mciteSetBstMidEndSepPunct{\mcitedefaultmidpunct}
{\mcitedefaultendpunct}{\mcitedefaultseppunct}\relax
\EndOfBibitem
\bibitem[Yoo \emph{et~al.}(2004)Yoo, Zeng, and Morris]{yooMeltingLinesModel2004}
S.~Yoo, X.~C. Zeng and J.~R. Morris, \emph{J. Chem. Phys.}, 2004, \textbf{120}, 1654--1656\relax
\mciteBstWouldAddEndPuncttrue
\mciteSetBstMidEndSepPunct{\mcitedefaultmidpunct}
{\mcitedefaultendpunct}{\mcitedefaultseppunct}\relax
\EndOfBibitem
\bibitem[Yoo \emph{et~al.}(2009)Yoo, Xantheas, and Zeng]{yooMeltingTemperatureBulk2009}
S.~Yoo, S.~S. Xantheas and X.~C. Zeng, \emph{Chem. Phys. Lett.}, 2009, \textbf{481}, 88--90\relax
\mciteBstWouldAddEndPuncttrue
\mciteSetBstMidEndSepPunct{\mcitedefaultmidpunct}
{\mcitedefaultendpunct}{\mcitedefaultseppunct}\relax
\EndOfBibitem
\bibitem[Yoo \emph{et~al.}(2009)Yoo, Zeng, and Xantheas]{yooPhaseDiagramWater2009}
S.~Yoo, X.~C. Zeng and S.~S. Xantheas, \emph{J. Chem. Phys.}, 2009, \textbf{130}, 221102\relax
\mciteBstWouldAddEndPuncttrue
\mciteSetBstMidEndSepPunct{\mcitedefaultmidpunct}
{\mcitedefaultendpunct}{\mcitedefaultseppunct}\relax
\EndOfBibitem
\bibitem[Vega \emph{et~al.}(2008)Vega, Sanz, Abascal, and Noya]{vegaDeterminationPhaseDiagrams2008}
C.~Vega, E.~Sanz, J.~L.~F. Abascal and E.~G. Noya, \emph{J. Phys. Condens. Matter}, 2008, \textbf{20}, 153101\relax
\mciteBstWouldAddEndPuncttrue
\mciteSetBstMidEndSepPunct{\mcitedefaultmidpunct}
{\mcitedefaultendpunct}{\mcitedefaultseppunct}\relax
\EndOfBibitem
\bibitem[MacDowell \emph{et~al.}(2004)MacDowell, Sanz, Vega, and Abascal]{macdowellCombinatorialEntropyPhase2004}
L.~G. MacDowell, E.~Sanz, C.~Vega and J.~L.~F. Abascal, \emph{J. Chem. Phys.}, 2004, \textbf{121}, 10145--10158\relax
\mciteBstWouldAddEndPuncttrue
\mciteSetBstMidEndSepPunct{\mcitedefaultmidpunct}
{\mcitedefaultendpunct}{\mcitedefaultseppunct}\relax
\EndOfBibitem
\end{mcitethebibliography}
\bibliographystyle{rsc}

\end{document}


\title{Electronic Supplementary Information: Active Sparse Bayesian Committee Machine Potential for Isothermal-Isobaric Molecular Dynamics Simulations}

\author{Soohaeng Yoo Willow}
\affiliation{Department of Energy Science, Sungkyunkwan University, Seobu-ro 2066, Suwon, 16419, Korea}

\author{Dong Geon Kim}
\affiliation{Department of Energy Science, Sungkyunkwan University, Seobu-ro 2066, Suwon, 16419, Korea}

\author{R. Sundheep}
\affiliation{Department of Energy Science, Sungkyunkwan University, Seobu-ro 2066, Suwon, 16419, Korea}

\author{Amir Hajibabaei}
\affiliation{Yusuf Hamied Department of Chemistry, University of Cambridge, Lensﬁeld Road, Cambridge, CB2 1EW, United Kingdom}

\author{Kwang S. Kim}
\affiliation{Center for Superfunctional Materials, Department of Chemistry, Ulsan National Institute of Science and Technology,  Ulsan 44919, Korea}  

\author{Chang Woo Myung}
\email{cwmyung@skku.edu}
\affiliation{Department of Energy Science, Sungkyunkwan University, Seobu-ro 2066, Suwon, 16419, Korea}

\date{\today}

\maketitle
\tableofcontents
\newpage

We provide additional supporting data as well as contextual information to the main text here. 
All input and output files are provided on \href{https://github.com/myung-group/Data_active_BCM.git}{GitHub}, which contains a Jupyter Notebook file that analyses the data.




\section{Active BCM MD simulation}

The goal of on-the-fly active learning MD simulations is to automatically construct both the dataset and the inducing set.
Using these sets, the weight parameter vector $\bm{w}$ is determined by fitting the predicted values of energy, forces, and virial pressures to their target values. During the MD simulations, we update the inducing set $z_P=\{\chi_j\}$ and dataset $X_P = \{\bm{R}_n\}$ according to the SGPR technique. The SGPR technique works as follows. 
\begin{enumerate}
    \item A new configuration $\bm{x}^*$ is generated from the MD simulation.
    \item The local chemical environments (LCEs) $\rho$ of atoms are generated.
    \item The kernel matrices for energy $\mathcal{K}(\rho_i, \chi_j)$, 
    forces $\dot{\mathcal{K}}_i^\mu (\bm{R}, \chi_j)$, 
    and virial pressures $\dot{\mathcal{K}}_i^\mu (\bm{R}, \chi_j) r_i^\nu$ are calculated as per Eq (6), (7), and (9).
    \item Energy, forces, and virial pressures are predicted using Eq. (11), (13), and (14) for the BCM potential.
    \item Check for new LCEs by calculating the covariance loss $\sigma^2 (\rho)$ between an LCE and the inducing sets of all local experts:
    \begin{eqnarray}
        \sigma^2(\rho) =  (1 - \mathbf{K}_{\rho m} \mathbf{K}_{mm}^{-1} \mathbf{K}_{\rho m}^T).
    \end{eqnarray}
    \item Update the inducing set when new LCEs are detected, with the condition $\sigma^2(\rho) > \sigma_\mathrm{cutoff}$. Here, $\sigma_\mathrm{cutoff} = 0.05$ .
    \item The updated inducing set $z=\{\chi_j\}$ changes the kernel matrices for energy, forces, and virial pressures, which in turn affects the predicted energy, forces, and virial pressures.
    If the change in the predicted energy $\Delta E$ is significant ($\Delta E > \Delta E_\mathrm{cutoff}$), we update the dataset by adding the new configuration $\bm{x}^*$ along with its energy, forces, and virial pressures. In our calculation, $\Delta E_\mathrm{cutoff} = 0.05$ eV.
\end{enumerate}

\begin{figure} [htp]
\centering
\includegraphics[width=16cm]{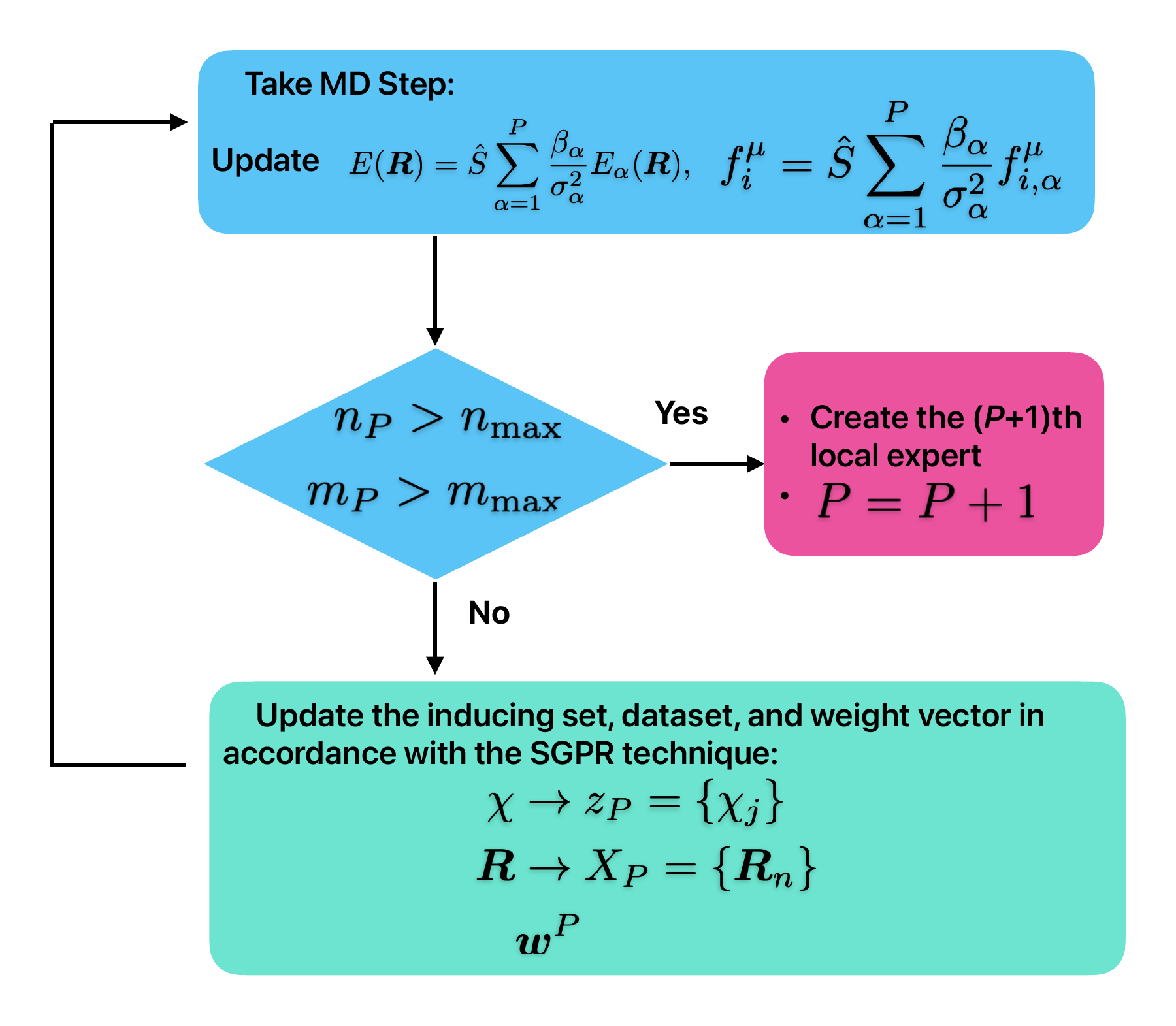}
\caption{On-the-fly active learning of Bayesian committee machine (BCM) potential during MD simulations. During each time-step of the MD simulation, the BCM potential calculates energy, forces, and stresses. 
If the number of data in $X_P = \{\bm{R}_n\}$ or inducing set $z_P = \{ \chi_j \}$ exceeds the threshold of a predefined kernel size (for example, $n_\mathrm{max} = 50$, $m_\mathrm{max} = 200$), a new local expert sparse Gaussian process regression potential model is trained.}

\end{figure}

\newpage 

\section{Liquid boron nitride $NVT$ MD simulation}

\begin{figure}[htp]
    \centering
    \includegraphics[width=8cm]{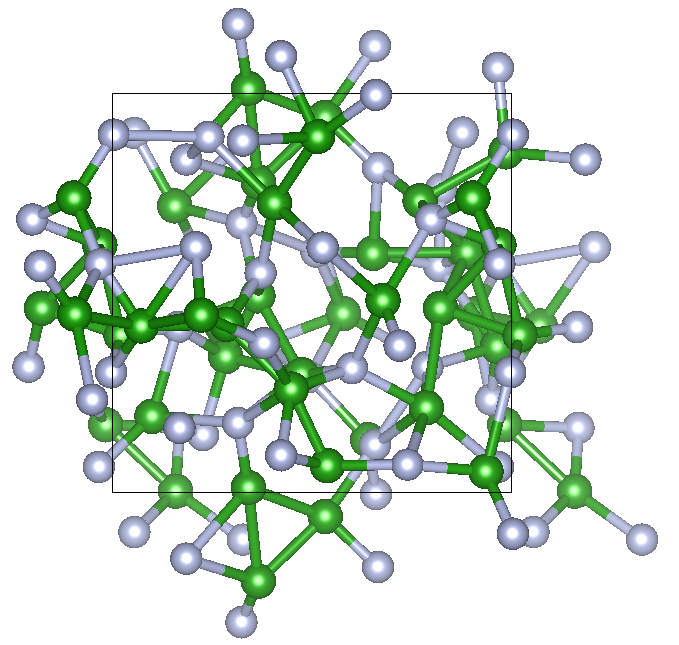}
    \caption{High-density liquid-phase BN Unit cell where green represents Boron atoms, and white represents Nitrogen atoms, respectively.}
    \label{fig:lbn}
\end{figure}

$NVT$ MD simulations on liquid boron nitride were carried out at $T = 7500$ K. The simulations were conducted on a unit cell (with a lattice parameter of $7.252 \text{\r{A}}$) consisting of 64 atoms as shown in Figure \ref{fig:lbn}.

\begin{figure}[htp]
    \centering
    \includegraphics[width=10cm]{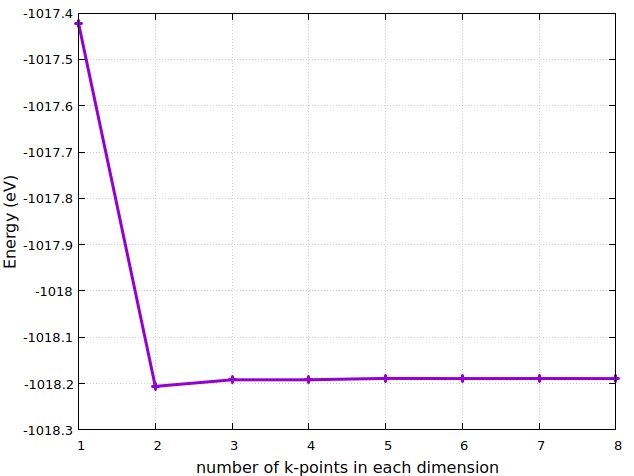}
    \caption{K-point grid convergence calculations for high density liquid phase of BN.}
    \label{fig:kpc}
\end{figure}

The MD simulations were carried out using the MD engine from the Atomic Simulation Environment (ASE) python library.\cite{larsen_atomic_2017} The simulations ran for 8 $ps$ using Nos{\'e}-Hoover dynamics with a target temperature of 7500 $K$, a time step of 1 $fs$ and time constant of 25 $fs$. For these ASE MD simulations, the forces were calculated using the Vienna Ab initio Simulation Package (VASP).\cite{PhysRevB.54.11169,kresseEfficiencyAbinitioTotal1996} 
Before running the MD simulations, we performed convergence calculations for cutoff energy and the k-point grid (Figure \ref{fig:kpc}) to achieve accurate DFT results. Based on these results, we employed an energy cutoff of 650 $eV$, and ($3 \times 3 \times 3$) Monkhorst-pack k-grid at the $r^2$SCAN-D4 functional level.\cite{r2scan, dftd4}

\newpage

\section{\ce{Li10Ge(PS6)2} $NpT$ MD simulation}
All DFT calculations were performed using the Vienna Ab-initio Simulation Package(VASP)\cite{PhysRevB.54.11169,kresseEfficiencyAbinitioTotal1996} with the projector augmented wave (PAW) method at the Perdew–Burke–Ernzerhof (PBE)\cite{perdewRationaleMixingExact1996} functional level to train the SGPR potential. The LGPS model, sourced from the Materials Project database (ID: mp-696128), consisted of 50 atoms (Li: 20, Ge: 2, P: 4, S: 24) within a tetragonal box measuring 8.8 Å $\times$ 8.8 Å $\times$ 12.7 Å. $NpT$ MD simulations  were  performed under the atmospheric pressure ($\sim$0.101 MPa) using a Nosé-Hoover thermostat and a Parrinello-Rahman barostat, with 1 $fs$ time step, 500 eV energy cut-off, and $(1 \times 1 \times 1)$ Monkhorst-pack k-point mesh.

We calculated the mean square displacement(MSD) as
\begin{equation}
   \mathrm{ MSD}(t)=\frac{1}{N} \sum_{i=1}^{N} \left\vert \Vec{r}_i(t)-\Vec{r}_i(0) \right\vert ^{2}
\end{equation}

where $N$ is the number of particles, $\Vec{r}_i(t)$ and $\Vec{r}_i(0)$ is the position of the $i$-th particle at time $t$ and reference position. The diffusivity ($D$) is given as 
\begin{equation}
    D=\lim_{t \rightarrow \infty} \frac{1}{2 d t} \langle \mathrm{MSD}(t) \rangle
\end{equation}
where $d$ is number of dimensions and $\langle \cdot \rangle$ indicates the ensemble average. We estimated the diffusion activation energy by assuming an Arrhenius temperature dependence
\begin{equation}
    D=D_0 e^{-E_a / k_B T}
\end{equation}


\begin{figure} [htp]
\centering
\includegraphics[width=6.5cm]{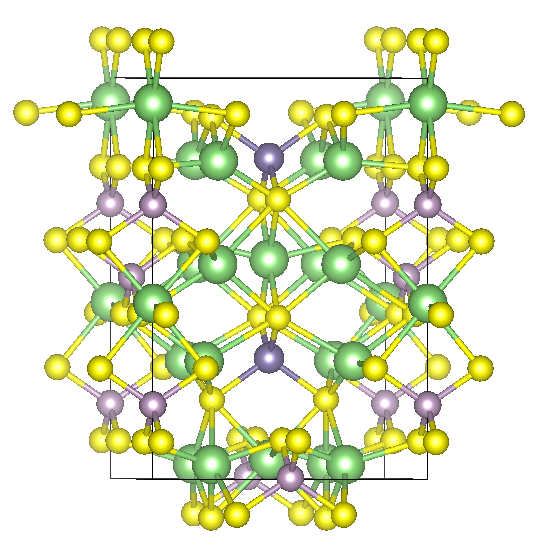}
\includegraphics[width=8.5cm]{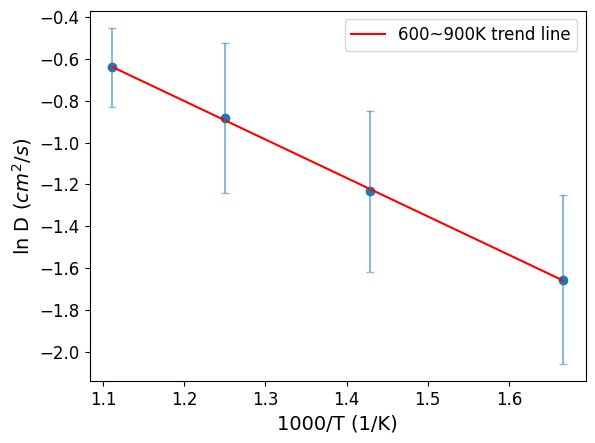}
\caption{The unit cell of LGPS (left) and Li-ion diffusivity ($cm^2/s$) with respect to temperature (right) (Li: green, S: yellow, P: purple, Ge: indigo). The red line represents the trend ranging from 600 K to 900 K. The error bars are estimated using the block average with block size of 10 $ps$.}
\label{fig:LGPS_D}
\end{figure}

\newpage

\section{Ice-Liquid $NpH$ MD Simulations}

\begin{table}[h]
\caption{\label{tab:S1}
Ice-liquid coexisting systems. Total number of water molecules in the simulation box ($N_{\mathrm{H}_2\mathrm{O}}$) and initial cell dimensions ($\Vec{a},\Vec{b},\Vec{c}$,$\alpha,\beta,\gamma$) of ice-liquid coexisting systems for $NpH$ MD simulations.}
\begin{tabular}{ |c |c| c | }
  \hline
  & $N_{\mathrm{H}_2\mathrm{O}}$ & ($|\Vec{a}|$, $|\Vec{b}|$, $|\Vec{c}|$, $\alpha$, $\beta$, $\gamma$) \\
  \hline
  Ice II - Liquid & 864 & (23.415, 23.415, 43.910, 84.826, 84.826, 113.100) \\
  Ice III- Liquid & 648 & (19.982, 20.349, 41.379, 90, 90, 90) \\
  Ice V - Liquid & 672 &  (18.553, 45.374, 20.833, 90, 109.215, 90) \\
  \hline
\end{tabular}
\end{table}

\begin{figure}[h!]
    \centering
    \includegraphics[width=0.8\textwidth]{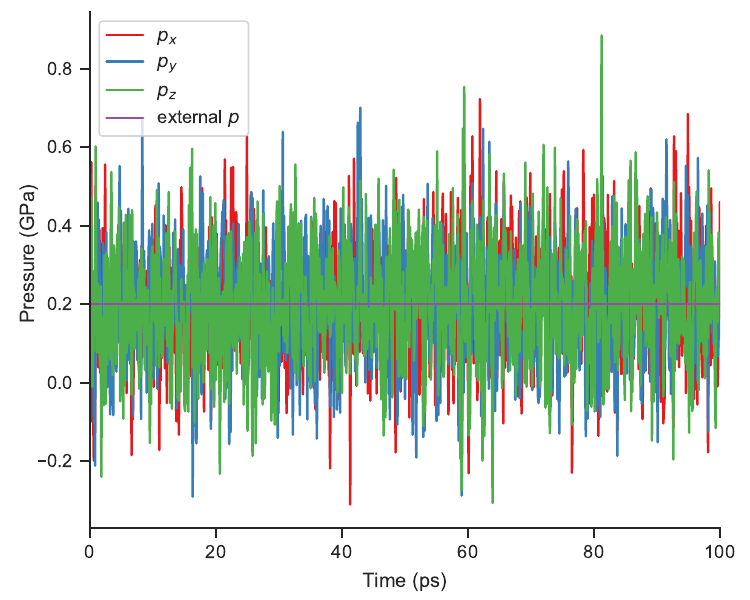}
    \caption{Instantaneous pressures $p_x$, $p_y$, and  $p_z$ values in anisotropic $NpH$ MD simulations of coexisting ice II-liquid system at pressure of $p=0.2$ GPa. Instantaneous pressures are adjusted to specified target pressure during $NpH$ MD simulations.  Note that anisotropic $NpH$ MD simulations show no significant pressure differences along different axes.}
    \label{fig:P}
\end{figure}

\newpage 

\begin{figure}[h!]
    \centering
    \includegraphics[width=1.0\textwidth]{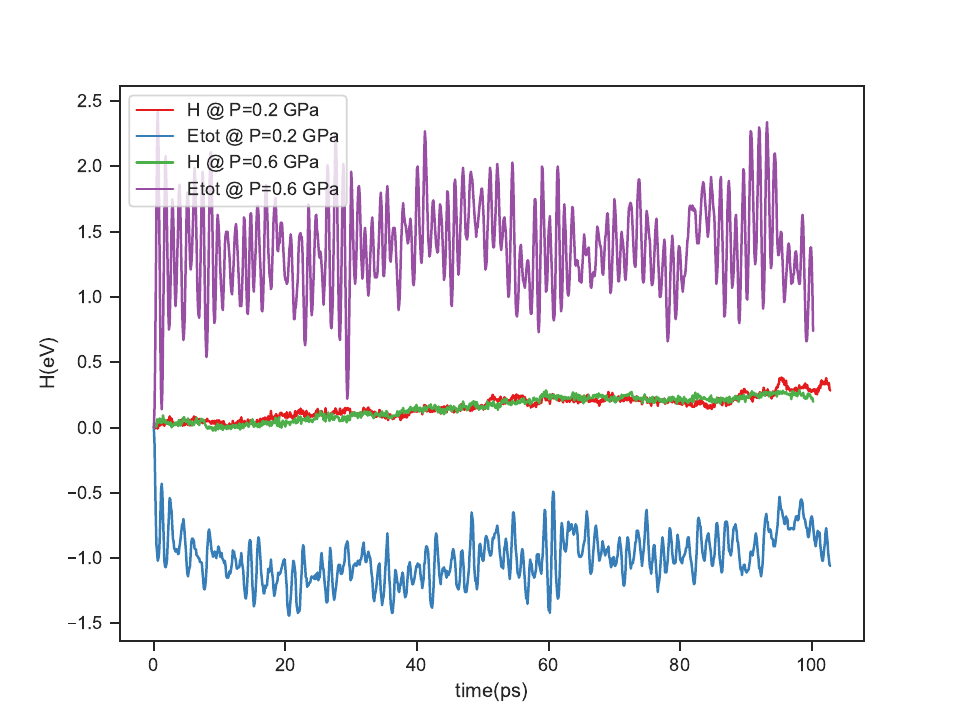}
    \caption{Enthalpy ($H$) and total energy ($E_\mathrm{tot}$) during $NpH$ MD simulations of coexisting ice II-liquid systems at pressures of $p=0.2$ and $0.6$ GPa. $H$ is conserved while the total energies fluctuate.}
    \label{fig:H}
\end{figure}

\newpage

\begin{figure}[h!]
    \centering
    \includegraphics[width=1.0\textwidth]{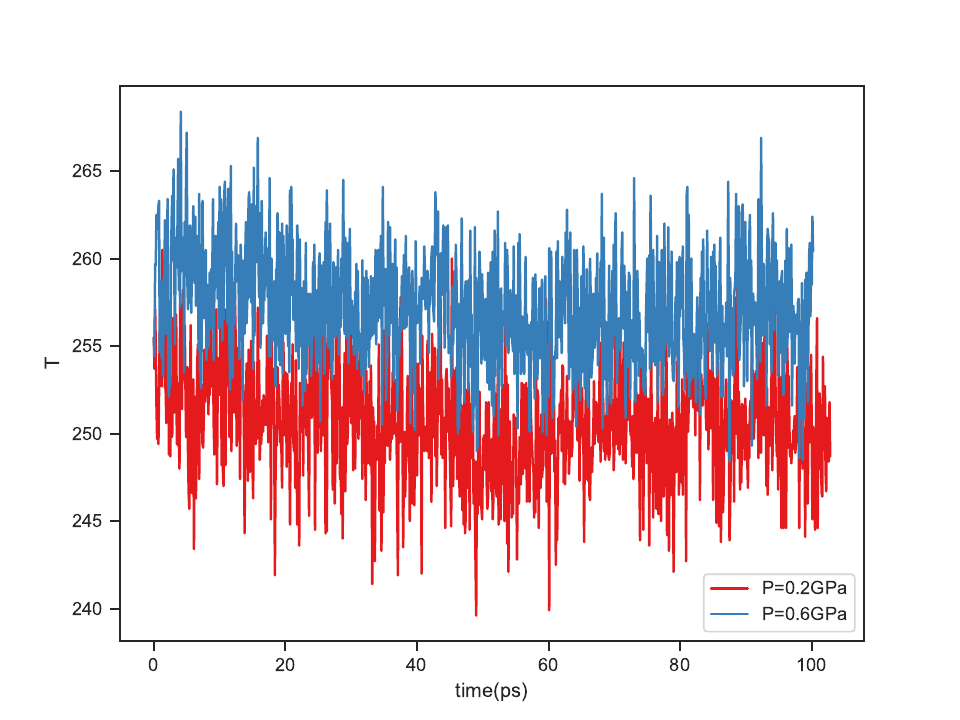}
    \caption{Instantaneous temperature values in $NpH$ MD simulations of coexisting ice II-liquid system at pressures of $p=0.2$ and $0.6$ GPa. 
    Instantaneous temperatures are tuned to approach melting temperature, ensuring that they meet the condition $\mu_\mathrm{ice} (p, T)_{T=T_m} = \mu_\mathrm{liq}(p,T)_{T=T_m}$. }
    \label{fig:T}
\end{figure}

\newpage 

\section{Melting points of ice-water for MB-pol}

\begin{figure}[h!]
    \centering
    \includegraphics[width=0.8\textwidth]{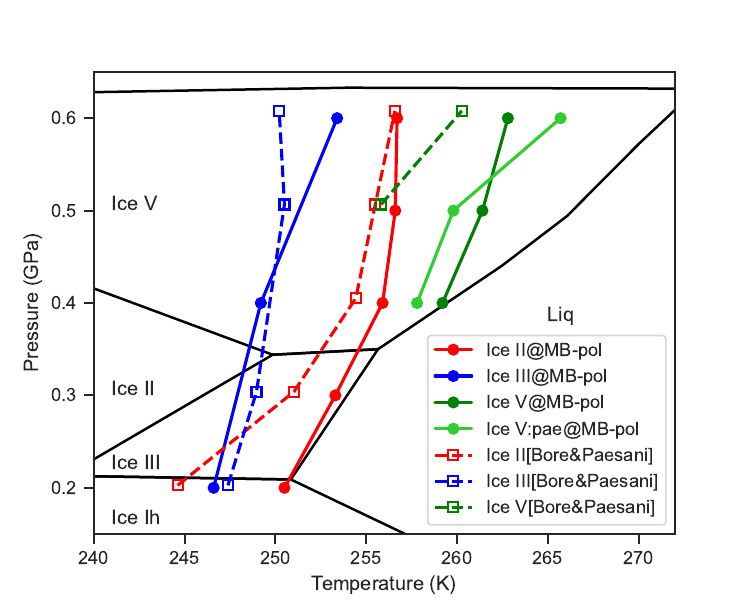}
    \caption{ 
    The melting points of ice-water for MB-pol in our study (shown with circle lines) compared to those in Paesani's study (indicated by square lines). 
    Melting points of Ice V:pae@MB-pol were obtained with the initial configuration that the Bore and Paesani used in their study.}
    \label{fig:ice_water_Tm}
\end{figure}

In Figure \ref{fig:ice_water_Tm}, we compared the melting points of ice-water using the MB-pol model in our study (shown with circle lines) with those from Bore and Paesani's study\cite{BorePaesani23} (indicated by square lines). 
While the melting lines of ice III are highly very similar, those of ice V are significant different. 

To understand why ice V behaves differently,
we aim to share the methods used in both studies. 
Bore and Paesani\cite{BorePaesani23} described their approach in their supplementary Note3: ``These calculations were carried out following a four-stage procedure ... (a) we determined the classical melting points from the chemical potential differences between liquid water and each ice polymorph using enhanced-coexistence simulations carried with the DNN@MB-pol potential, (b) starting from the classical melting points obtained with DNN@MB-pol, we used thermodynamic perturbation theory to determine the classical melting points of MB-pol, (c) ...''

\begin{eqnarray}
    \exp \left[-\beta \Delta G (p, T)  \right] & = & \langle \exp [ -\beta \Delta U] \rangle_{p,T,\mathrm{DNN@MB-pol}}   \label{eq:SI_TI} \\
    \Delta U & = & {U_\mathrm{MB-pol}(R_\mathrm{DNN@MB-pol}) - U_\mathrm{DNN@MB-pol}(R_\mathrm{DNN@MB-pol})} \label{eq:SI_dU} \\
    \Delta \mu & = & \mu_\mathrm{MB-pol} - \mu_\mathrm{DNN@MB-pol} = \frac{\Delta G}{N_\mathrm{H_2O}} \label{eq:SI_dmu}
\end{eqnarray}

To calculate the chemical potentials $\mu_\mathrm{MB-pol}$ of the MB-pol water model in Eq. (\ref{eq:SI_dmu}) using Eq. (\ref{eq:SI_TI}), 
they utilized the coordinates $\{ R_\mathrm{DNN@MB-pol} \}$ sampled within the DNN@MB-pol potential model 
to compute the potential energy $U_\mathrm{MB-pol}$ in Eq. (\ref{eq:SI_dU}).
Eq. (\ref{eq:SI_TI}) represents the mathematical formulation of the free energy perturbation (FEP) identity introduced by Zwanzig in 1954\cite{zwanzigHighTemperatureEquation1954} rather than thermodynamic perturbation theory. 
Generally, 
free energy perturbation is only valid when the coordinates sampled from DNN@MB-pol closely match those from MB-pol. 
Hence, the accuracy of the free energy difference calculated using Eq. (\ref{eq:SI_TI}) strongly relies on how much two thermodynamic states overlap in configuration space. To mitigate this overlap requirement, a series of intermediate thermodynamic states are introduced between two thermodynamic end states $U_\mathrm{MB-pol}$ and $U_\mathrm{DNN@MB-pol}$ as $(1-\lambda)U_\mathrm{DNN@MB-pol} + \lambda U_\mathrm{MB-pol}$. 
Because Bore and Paesani\cite{BorePaesani23} did not performed MD simulations with all intermediate and the final thermodynamic states, their estimated potential energies were inaccurate due to neglecting the requirement of the overlap between adjacent thermodynamic states.

In our study, on the other hand,
we sampled all configurations $\{R_\mathrm{MB-pol} \}$ from $NpH$ MD simulations with the MB-pol water model when we estimate the melting points for MB-pol. 
Similarly, the melting points for the BCM-MLP model were estimated with all configurations $\{ R_\mathrm{BCM-MLP}\}$ sampled from $NpH$ MD simulations with the BCM-MLP model. In short, we directly estimated the melting points, which rely on their specific thermodynamic states and potential energy surface.

In summary, the main reason why our melting points and theirs don't match up, especially melting points of ice V,
is because we used different sets of configurations.
To estimate the melting points for the MB-pol model, they used the configurations from DNN@MB-pol, while we used ones from MB-pol directly.
We predict that if they had calculated the melting points using the MB-pol potential instead of using DNN@MB-pol potential, 
the melting points of ice V would be much more similar.

\newpage

\bibliography{SGPR}
\bibliographystyle{rsc}